\documentclass[reviewcopy]{elsarticle}

\usepackage[reviewcopy]{adndt}
\usepackage{longtable}

\usepackage{amsmath}

\usepackage{amssymb}

\biboptions{square,sort&compress}
\bibpunct[]{[}{]}{,}{n}{}{;}
\begin{document}

\begin{frontmatter}

\journal{Atomic Data and Nuclear Data Tables}

\title{ \normalfont\textsc{Ground State and Saddle Point: masses and deformations for even-even superheavy nuclei with $98\leq Z \leq 126$
and $134\leq N \leq 192$ }}

  \author[One]{M. Kowal\corref{cor1}}
  \ead{E-mail: m.kowal@fuw.edu.pl}

  \author[Two]{P. Jachimowicz}

  \author[One]{J. Skalski}

  \cortext[cor1]{Corresponding author.}

  \address[One]{National Centre for Nuclear Research , Ho\.za 69,
PL-00-681 Warsaw, Poland}

  \address[Two]{Institute of Physics,
University of Zielona G\'{o}ra, Szafrana 4a, 65516 Zielona
G\'{o}ra, Poland}

\date{\today} 

\begin{abstract}
 We determine ground-state and saddle-point
 shapes and masses of even-even superheavy nuclei in the
range of proton numbers $98\leq Z \leq 126$ and neutron numbers
$134\leq N \leq 192$. Our study is performed within the
 microscopic-macroscopic method. The Strutinsky shell and pairing correction
 is calculated for the deformed Woods-Saxon single-particle potential
 and the Yukawa-plus-exponential energy is taken as a smooth part.
 We use parameters of the model that were fitted previously to
 this region of nuclei.
 A high-dimensional deformation space, including nonaxial and
 reflection-asymmetric shapes, is used in the search for saddle points.
 Both ground-state and saddle-point shapes are found with the aid of the
 minimization procedure, with dynamical programming technique of search for
  saddle points.
 The results are collected in two tables. Calculated ground-state
 mass-excess, $Q_{\alpha}$ energies, total and macroscopic
 energies normalized to the macroscopic energy at the spherical shape,
 shell corrections (including pairing) and deformations
are given for each nucleus in the table one. The second table gives
 the same properties, but at the saddle-point configuration.
 The obtained results are discussed and
 compared with available experimental data for
 alpha-decay energies ($Q_{\alpha}$) and ground-state masses.
\end{abstract}

\end{frontmatter}




\newpage

\tableofcontents
\listofDtables
\listofDfigures
\vskip5pc


\section{Introduction}

The uncharted region of the $Z$-$N$ plane can answer many questions of fundamental importance for science:
How many neutrons can be bound in a nucleus? What are the unique properties of
 short-lived nuclei having extreme values of $N/Z$? What are the properties of
 effective nuclear interactions in the environment different from that in
 stable nuclei?
 There is also an unsettled question of what can be the largest possible
 atomic number $Z$ of an atomic nucleus.
 The recent experiments in Dubna claim the existence
 of a $Z=114,115,116,117,118$ system \cite{O1,O2,O3,O4,O5,117,118,120,O6,O7,O8}, with
 confirmation of hot fusion cross-sections coming from GSI \cite{GSI} and
 LBL Berkeley \cite{LBL}. For lighter elements: $Z=107,108,109,110,111,112$ successful syntheses were
done at the GSI laboratory \cite{m1,m2,m3,S1,S2,S3,S4,S5,S6,S7}.

The presented tables contain ground state (g.s.) and saddle point (s.p.)
  properties of even-even superheavy nuclei that were obtained within the
 microscopic-macroscopic method in a multidimensional deformation space.
 This conceptually simple method is very convenient for determining many
 nuclear properties with relatively high accuracy, unattainable
 by other models, particularly in the area of heaviest elements.
 It makes possible global calculations of fission properties in this area of
 the atomic nuclei.
 Such systematic predictions relate to the questions specified above
 and contribute to a view on this exotic nuclear domain.
  At the same time, they can help in the interpretation of present
 experiments and inspire efforts towards new ones.

 Atomic masses (binding energies) have been experimentally determined for
 nearly 2300 nuclei \cite{Audi2003}.
 For almost 800 nuclei mass is measured more precisely than to 5 keV,
for other ~1000 nuclei - with the accuracy of 5-50 keV.
 The binding energy determines energy available for nuclear reactions
 and decays (and thus the creation of elements by stellar nucleosynthesis),
and holds the key to the fundamental question of how the heavy elements
 came to existence.
The best available theoretical global mass calculations predict the mass
 values with an approximate deviation 500-800 keV.
Ground state masses calculated within the Hartree-Fock-Bogoliubov (HFB) method
 \cite{Gor1,Gor2} fitted to the fission data through
adjustment of a vibrational term in the phenomenological collective correction
 have the r.m.s deviation equal 0.729 MeV.
Macroscopic-microscopic global calculations of nuclear ground state masses
 made by P. Moller and co-workers \cite{mol95}
give the r.m.s error 0.669 MeV for nuclei ranging from Oxygen to Hassium and
 0.448 MeV in the case of nuclei above $N=65$.
Phenomenological formula with the 10 free parameters given by Duflo and Zuker
 \cite{Duflo1,Duflo2} gives mass estimates  with the 0.574 MeV r.m.s error.
So, nuclear masses are presently measured
   at least ten times more precisely than they can be calculated.

Although theoretical predictions sometimes differ quantitatively,
they consistently predict prolate deformed superheavy nuclei with $Z=100$-$112$,
which is confirmed experimentally for nuclei around $^{254}$No \cite{No},
and spherical or oblate deformed systems with $Z=114$
 and $N=174$-$184$ \cite{Cwiok83,Leander,Boning,Sob87,611,CHN}.
There are relatively many published predictions of the ground
state properties
 of heaviest elements  \cite{Gor1,Gor2,Duflo1,Duflo2,mol95,BARAN05}.
 Much rarer are analogous calculations at the saddle points, although
 they are necessary to estimate the cross sections (survival probabilities)
 for the synthesis of SHN \cite{Wil2}. This work provides the necessary
 theoretical saddle point data.

 As the principle of the macroscopic -microscopic method is well known,
 only a brief description of it is given in Sect. 2,
 specifying involved quantities, their computation and the adopted
 values of parameters.
 In Sect. 3 we describe a variety of considered nuclear shapes which is
 an important ingredient of the model. The method for saddle point search
 is described in section 4, numerical tests applied to certify
 the results are mentioned in Sect. 5. Results and discussion are given in
 Sect. 6, a short summary in Sect. 7.

\section{Method}

The total nuclear binding energy ($E$), which depends on
 the proton number Z, and neutron number N and the nuclear shape,
 can be written as a sum of a macroscopic ($E_{mac}$) and a microscopic
($E_{mic}$) energy:

\begin{eqnarray}
E(def,Z,N) = E_{mic}(def,Z,N) + E_{mac}(def,Z,N).
\end{eqnarray}

\subsection{Macroscopic energy}

 The macroscopic part of atomic mass is a sum of masses of atomic
 constituents and the macroscopic energy.  As $E_{mac}$, it is a smooth
function of proton and neutron number. In our analysis, it is
taken in the liquid-drop form \cite{kra79,mun01a}:

\begin{eqnarray} \label{mmacr}
 M_{\rm macr}(Z,N,\beta_{\lambda}^0)&=&M_{\rm H} Z+M_{\rm n} N
-a_{\rm v}(1-\kappa_{\rm v} I^2)A +a_{\rm s}(1-\kappa_{\rm s}
I^2)A^{2/3}B_S(\{\beta_{\lambda}^0\}) \nonumber\\
&& + a_0 A^0 +c_1 Z^2 A^{-1/3} B_C(\{\beta_{\lambda}^0\}) -c_4 Z^{4/3}
A^{-1/3} \nonumber\\
&&+ f(k_{\rm F} r_{\rm p})Z^2 A^{-1} -c_{\rm a}(N-Z)
-a_{\rm el}Z^{2.39},
\end{eqnarray}
where $M_{\rm H}$ is mass of the hydrogen atom, $M_{\rm n}$ is
mass of neutron, $I=(N-Z)/A$ is the relative neutron excess,
$A=Z+N$ is the mass number of a nucleus. The functions
$B_{S}(\beta_{\lambda})$ and $B_{C}(\beta_{\lambda})$ describe the
dependence of the surface and Coulomb energies, respectively, on
 deformations $\beta_{\lambda}$, and $\beta_{\lambda}^0$ are the
values of these deformations at equilibrium. We adopted these
functions in the form given by the Yukawa-plus-exponential model
formulated by Krappe and Nix \cite{kra79}. They read
\cite{mol81+,mol95}:
\begin{equation}
B_S = \frac{A^{-2/3}}{8\pi^2r_0^2a^4}\int\int_V \left(
2-\frac{r_{12}}{a} \right) \frac{e^{-r_{12}/a}} {r_{12}/a} d^3r_1
d^3r_2,
\end{equation}
\begin{equation}
B_C = \frac{15}{32\pi^2} \frac{A^{-5/3}}{r_0^5}\int\int_V
\frac{1}{r_{12}} \left[ 1- \left(
1+\frac{1}{2}\frac{r_{12}}{a_{\rm den}} \right) e^{-r_{12}/a_{\rm
den}} \right] d^3r_1 d^3r_2,
\end{equation}
where $r_{12}$=$|\overrightarrow{r_1}-\overrightarrow{r_2}|$ with
$\overrightarrow{r_1}$ and $\overrightarrow{r_2}$ describing the
positions of two interacting volume elements, $a$ is the range of
the Yukawa interaction on which the model is based, $a_{\rm den}$
is the range of the Yukawa function used to generate nuclear
charge distribution. The functions are normalized in such a way
that they are equal 1 for a spherical nucleus in the limit
case of $a$=0 (for $B_S$) and $a_{\rm den}$=0 (for $B_C$),
corresponding to the traditional liquid-drop model with a sharp
surface. The integrations are over the volume of a nucleus.
 After turning them into surface integrals, $B_S$ and $B_C$ were calculated
 by using a four-fold (or three-fold, for axial symmetry) 64-point
 Gaussian quadrature.

The quantities $c_1$ and $c_4$ appearing in the Coulomb energy and the
Coulomb exchange correction, respectively, are
\begin{equation}
c_1 = \frac{3}{5} \frac{e^2}{r_0},
\qquad c_4 = \frac{5}{4} \left( \frac{3}{2\pi} \right)^{2/3} c_1,
\end{equation}
where $e$ is the elementary electric charge and $r_0$ is the
nuclear-radius parameter.
The quantity $f(k_{\rm F} r_{\rm p})$
appearing in the proton form-factor correction to the Coulomb
energy in Eq.~(\ref{mmacr}) has the form
\begin{equation}
f(k_{\rm F} r_{\rm p}) = -\frac{1}{8} \frac{e^2r_{\rm p}^2}{r_0^3}
\left[
\frac{145}{48} - \frac{327}{2880}(k_{\rm F} r_{\rm p})^2 +
\frac{1527}{1\,209\,600}(k_{\rm F} r_{\rm p})^4
\right],
\end{equation}
where the Fermi wave number is
\begin{equation}
k_{\rm F} = \left( \frac{9\pi Z}{4A} \right)^{1/3} r_0^{-1},
\end{equation}
and $r_{\rm p}$ is the proton root-mean-square radius. The last
term in Eq.~(\ref{mmacr}) describes the binding energy of
electrons and $a_{\rm v}$, $\kappa_{\rm v}$, $a_{\rm s}$,
$\kappa_{\rm s}$, $a_0$, $c_{\rm a}$ are adjustable parameters.
Thus, only two of these parameters ($a_{\rm s}$ and $\kappa_{\rm
s}$) appear at the term, which depends on deformation. The four
remaining parameters stand at the terms independent of the shape
of a nucleus.

  The macroscopic part of mass, Eq.~(\ref{mmacr}), is used the same
as in \cite{mun01a}, except that three of its adjustable
parameters: $a_{\rm v}$, $\kappa_{\rm v}$ and $a_0$ were fitted to
experimental masses of even-even heaviest nuclei with
$Z$$\geqslant$84. The result was
\begin{equation}\label{det_pars1}
a_{\rm v} = 16.0643, \qquad \kappa_{\rm v} = 1.9261, \qquad a_0 =
17.926.
\end{equation}
Following the authors of \cite{mun01a}, we omit here the two terms
considered in \cite{mol81+}: charge-asymmetry term $c_{\rm
a}(N-Z)$ and Wigner term (characterized by a coefficient $W$),
 as they do not significantly change the quality of the description of masses
of heaviest nuclei. The values of other parameters are adopted after \cite{mol81+}:
\begin{equation}\label{det_pars3}
a_{\rm s} = 21.13 \; {\rm MeV}, \qquad \kappa_{\rm s}=2.30,
\end{equation}
\begin{eqnarray}\label{det_pars4}
&&a = 0.68 \; {\rm fm}, \qquad a_{\rm den}=0.70\;{\rm fm}, \qquad
r_0=1.16\;{\rm fm}, \nonumber \\
&&r_{\rm p}=0.80\;{\rm fm}, \qquad a_{\rm el} = 1.433\cdot
10^{-5}\;{\rm MeV}.
\end{eqnarray}

\subsection{Microscopic energy}

The Strutinski shell correction \cite{str67,str68}, based on the
deformed Woods-Saxon single-particle potential, is taken for the
microscopic part:
\begin{eqnarray}  \label{mic}
 E_{mic}(def,Z,N) &=&  E^{\rm sh}_{\rm corr}(def,Z,N)\nonumber\\
                  &+&  E^{\rm pair}_{\rm corr}(def,Z,N),
\end{eqnarray}
where $E^{\rm sh}_{\rm corr}$ and $E^{\rm pair}_{\rm corr}$ are
the shell and pairing corrections, respectively.

 \subsubsection{ Woods-Saxon potential}

The Woods-Saxon potential $V_{\rm WS}$ has the following form:
\begin{equation}\label{ws_pot}
V_{\rm WS}(\vec{r})=-\frac{V}{1+e^{d(\vec{r},{\rm def})/a_{\rm
ws}}},
\end{equation}
where $V$ is the depth of the potential, $d(\vec{r},{\rm def})$ is
the distance from the point $\vec{r}$ to the surface of the
nucleus, $a_{\rm ws}$ is the diffuseness of the nuclear surface.
The symbol {\it def} stands for deformation which defines the
 nuclear surface (see Sect. 3). The depth of the potential is
\begin{equation}\label{ws_pot_depth}
V = V_0 (1 \pm \kappa I),
\end{equation}
where $I=(N-Z)/A$ is the relative neutron excess and $V_0$ and
$\kappa$ are adjustable parameters. The sign ($+$) is for protons
and ($-$) for neutrons.

In the case of spherical shape, the potential is
\begin{equation}\label{ws_pot_sph}
V_{\rm WS}(\vec{r})=-\frac{V}{1+e^{(r-R_0)/a_{\rm ws}}},
\end{equation}
where $R_0 = r_0 A^{1/3}$.

The full microscopic potential has the form (e.g. \cite{cwi87}):
\begin{equation}\label{micr_pot}
V_{\rm micr}=V_{\rm WS}+\lambda\left(\frac{\hbar}{2mc}\right)^2
\left(\frac{A}{A-1}\right)^2 \left( \nabla V^{\rm s.o.}_{\rm WS
}\right) \cdot \left(\vec{\sigma}\times\vec{p}/\hbar\right)+V_{\rm
c},
\end{equation}
where the second term is the spin-orbit potential and the third
term is the Coulomb potential, which has the following form:
\begin{equation}\label{coul_pot}
V_{\rm c}(\vec{r})=\rho_{\rm
c}\int\frac{d^3r'}{|\vec{r}-\vec{r'}|},
\end{equation}
where $\rho_{\rm c}=3(Z-1)e/(4\pi R_0^3)$ is the uniform density
and the
 integration extends over the volume enclosed by the nuclear surface.

 Here we use the ''universal'' set of parameters of the potential given in
\cite{cwi87}
\begin{equation}
\begin{array}{lll}
r_0 = 1.275 \;{\rm fm},  & (r_0)_{\rm so} = 1.32 \;{\rm fm}, &
\lambda = 36.0 \; {\rm for \;protons},\nonumber\\[3mm]
r_0 = 1.347 \;{\rm fm},  & (r_0)_{\rm so} = 1.31 \;{\rm fm}, &
\lambda = 35.0  \; {\rm for\;
neutrons},\\[3mm]
V_0 = 49.6  \;{\rm MeV}, & a_{\rm ws} = 0.70 \;{\rm fm}, & \kappa
= 0.86,
\end{array}
\end{equation}
\vspace{3mm}

\noindent where $r_0$ and $(r_0)_{\rm so}$ are the radius
parameters for the central and spin-orbit parts of the potential,
respectively.

 The single-particle potential is diagonalized in the
deformed-oscillator basis. The $n_{p}=450$ lowest proton levels
and $n_{n}=550$ lowest neutron levels from the $N_{max}=19$ lowest
shells of the deformed harmonic oscillator are taken into account
in the diagonalization procedure. We have determined the single -
particle spectra for every investigated nucleus. These
calculations therefore do not include any scaling relation to the
\emph{central} nucleus. A standard value of
$\hbar\omega_{0}=41/A^{1/3}$ MeV is taken for the oscillator
energy.


\subsubsection{ Shell correction}

    The shell correction energy is calculated as proposed by
Strutinski \cite{str67,str68}:
%
\begin{equation}\label{Esh_eq}
E^{\rm sh}_{\rm corr}=E_{\rm micro}-\widetilde{E}_{\rm micro},
\end{equation}
where $E_{\rm micro}$ is the sum of single-particle energies over
all occupied energy levels,
\begin{equation}\label{sp_sum1}
E_{\rm micro}
=\sum_{\nu_{\rm occ}} \varepsilon_\nu
=\int^{\varepsilon_{\rm F}}_{-\infty} \rho
(\varepsilon )\varepsilon d \varepsilon ,
\end{equation}
and
\begin{equation}\label{sp_dens}
\rho(\varepsilon)=\sum_\nu \delta(\varepsilon-\varepsilon_\nu)
\end{equation}
is the density of the single-particle levels per energy unit,
 $\varepsilon_{\rm F}$ is the Fermi energy and $\varepsilon_\nu$
is the energy of a single-particle level $\nu$.

The "smooth" microscopic energy
$\widetilde{E}_{\rm micro}$ is defined by means of the "smooth" density of
the single-particle levels $\widetilde{\rho}(\varepsilon)$:
\begin{equation}\label{sm_sp_en}
\widetilde{E}_{\rm micro}=\int^{\widetilde{\varepsilon}_{\rm
F}}_{-\infty} \widetilde{\rho} (\varepsilon )\varepsilon d
\varepsilon .
\end{equation}
The value of $\widetilde{\varepsilon}_{\rm F}$,
found from the following condition for the particle number $N$:
\begin{equation}\label{sh_cor1}
N=\int^{\varepsilon_{\rm F}}_{-\infty} \rho (\varepsilon ) d
\varepsilon=\int^{\widetilde{\varepsilon}_{\rm
F}}_{-\infty}\widetilde{\rho} (\varepsilon) d\varepsilon ,
\end{equation}
 is in general different from the Fermi energy $\varepsilon_{\rm F}$.

The "smooth" density, appearing in (\ref{sm_sp_en}) and
(\ref{sh_cor1}), is obtained as
\begin{equation}\label{smooth_den_def}
\widetilde{\rho}(\varepsilon)= \frac{1}{\gamma}
\int^{\infty}_{-\infty} \rho (\varepsilon^{\prime} )
f_p\left(\frac{\varepsilon^{\prime}-\varepsilon}{\gamma}\right) d
\varepsilon^{\prime},
\end{equation}
where $f_p$ is a folding function of the
Gaussian type, taken as the formal expansion of
 the $\delta$-function, truncated to
 the first $2p$ terms:
\begin{equation}\label{delta_expand_approx}
 f_p(x) = \frac{1}{\sqrt{\pi}} \sum_{n=0}^{2p} C_n H_n(x)
e^{-x^2},
\end{equation}
 with
\begin{equation}\label{C_n}
C_n= \frac{1}{2^n n!} H_n(0) = \left\{
\begin{array}{ll}
\frac{(-1)^\frac{n}{2}}{2^n(\frac{n}{2})!} & \textrm{for even $n$} \\
0 &  \textrm{for odd $n$.} \\
\end{array} \right.
\end{equation}
 The width $\gamma$ is of the order of shell energy gaps.
 Using $f_p$ one obtains the averaged density $\widetilde{\rho}$:
\begin{equation}\label{smooth_dens}
\widetilde{\rho}(\varepsilon
)=\frac{1}{\gamma\sqrt{\pi}}\sum_{\nu=1} e^{-u_\nu^2}
\sum_{n=0}^{2p}C_n H_n(u_\nu),
\end{equation}
where $u_\nu=(\varepsilon-\varepsilon_\nu)/\gamma$.

 The energy (\ref{sm_sp_en}) in general
depends on the parameters $\gamma$ and $p$. The method is
 meaningful, if there is a certain interval of $\gamma$ and
corresponding $p$, for which the energy does not practically
depend on them (so called "plateau condition").
Here, we use $\gamma=1.2 \hbar\omega_{0}$ for the Strutinski smearing parameter
 and a sixth-order correction polynomial for $f_p$.

 \subsubsection{ Pairing correlations}
 In this work, pairing is included within the
 Bardeen-Cooper-Schrieffer (BCS) theory \cite{bar57}.
 We assume a constant matrix element $G$ of the (short-range)
 monopole pairing interaction. The hamiltonian of a system of
  nucleons, separately for neutrons and protons, may be written as:
\begin{equation}\label{bcs_ham}
  H = \sum_{\nu}\varepsilon_\nu a_{\nu}^{+} a_{\nu} -
  G \sum_{\nu,\nu^{'}>0}  a_{\nu}^+ a_{\nu^{'}}^{+}
  a_{\bar{\nu}^{'}}  a_{\bar{\nu}},
\end{equation}
where $\varepsilon_{\nu}$ denotes the energy of a single-particle
state $\nu$. Each state $\nu$ has its time-reversal-conjugate
 $\bar{\nu}$ with the same energy (Kramers degeneration).

 As the BCS wave function is a
superposition of components with different numbers of particles,
 one requires that the
expectation value of the particle number has a definite value $N$:
\begin{equation}\label{bcs_N}
  \left<\hat{N}\right>=2\sum_{\nu>0}v_{\nu}^2 =
  N.
\end{equation}

The occupation numbers are given by
\begin{equation}\label{bcs_vnu}
    v_{\nu}^2=\frac{1}{2}\left[ 1-(\varepsilon_{\nu}-\lambda)/E_{\nu} \right],
\end{equation}
where
\begin{equation}\label{bcs_E_quasi}
    E_{\nu}=\sqrt{\left(\varepsilon_{\nu}-\lambda\right)^2+\Delta^2}.
\end{equation}
The parameters $\lambda$ and $\Delta$ are solutions of the system
of two equations, for the average particle number and the pairing gap:
\begin{equation}\label{bcs_gap}
\qquad\quad  N = \sum_{\nu>0} \left[1-
  \frac{\varepsilon_{\nu}-\lambda}{\sqrt{\left(\varepsilon_{\nu}-\lambda\right)^2+\Delta^2}} \right]
\end{equation}
\begin{equation}\label{bcs_gap1}
  \frac{2}{G} = \sum_{\nu>0}
  \frac{1}{\sqrt{\left(\varepsilon_{\nu}-\lambda\right)^2+\Delta^2}}.
\end{equation}

For the energy of the system in the BCS state, one gets:
\begin{equation}\label{bcs_ener}
  E_{\rm BCS} =
  2\sum_{\nu>0} \varepsilon_{\nu} v_{\nu}^2
  - \frac{\Delta^2}{G}
  - G\sum_{\nu>0} v_{\nu}^4.
\end{equation}

\subsubsection{ Pairing correction}
Pairing correction energy $E^{\rm pair}_{\rm corr}$ is usually
constructed in analogy to the shell correction energy $E^{\rm
sh}_{\rm corr}$,
\begin{equation}\label{bcs_E_pair}
E^{\rm pair}_{\rm corr} = E_{\rm pair} - \widetilde{E}_{\rm pair},
\end{equation}
where $E_{\rm pair}$ is the pairing energy corresponding to real
single-particle level distribution $\rho(\varepsilon)$,
Eq.~(\ref{sp_dens}), and $\widetilde{E}_{\rm pair}$ is this
energy for the smoothed s.p. level distribution,
$\widetilde{\rho}(\varepsilon)$, Eq.~(\ref{smooth_den_def}).

The $E_{\rm pair}$ is
\begin{equation}
E_{\rm pair} = E_{\rm BCS} - E_{\rm BCS}^{\Delta=0},
\end{equation}
where $E_{\rm BCS}^{\Delta=0}$ is the $E_{\rm BCS}$ energy in the
limit of disappearing pairing correlations ($\Delta = 0$). Thus,
using Eq.~(\ref{bcs_ener}),
\begin{equation}\label{bcs_ener_limit}
E_{\rm BCS}^{\Delta=0} =
  2\sum_{\nu=1}^{N/2} \varepsilon_{\nu}
  - \frac{GN}{2}.
\end{equation}
because for $\Delta=0$, the probability $v_{\nu}^2$ of
the occupation of any state $\nu$ is either 0 or 1.

 The smoothed pairing energy term is included in a schematic form, resulting
  from a model with a constant level density of pairs
 (doubly degenerate levels) ${\bar {\rho}}$, taken equal
 to $\widetilde{\rho}(\widetilde{\varepsilon_{\rm F}})/2$
 \begin{equation}
\widetilde{E}_{\rm pair} = -\frac{N_p^2}{{\bar \rho}}
(\sqrt{1+x^2}-1)
  +  \frac{{\bar G}N_p x}{2} \arctan(1/x) .
 \end{equation}
  In the above expression, $x={\bar \rho} {\bar \Delta}/{N_p}$, $N_p=N/2$ is
  a number of pairs, ${\bar \Delta}$ is an average value of the pairing gap
  in the neighbourhood of a studied nucleus, related to the average pairing
  strength ${\bar G}$ via the BCS formula for a constant level density
 \begin{equation}
  \frac{1}{{\bar G}{\bar \rho}} = \ln\left(\frac{\sqrt{1+x^2}+1}{x}\right) .
 \end{equation}
 The values of ${\bar \Delta}$ are taken from the fit \cite{MN88}
 \begin{equation}
   {\bar \Delta} = \frac{5.72}{N^{1/3}}\exp(-0.119 I -7.89 I^2)
 \end{equation}
  for neutrons and
 \begin{equation}
   {\bar \Delta} = \frac{5.72}{Z^{1/3}}\exp(0.119 I -7.89 I^2) ,
 \end{equation}
  for protons, with $I=(N-Z)/A$.
 The smoothed pairing energy term calculated in this way
 shows nearly no deformation dependence, for example, it varies by about
  50 keV  over the whole deformation range in actinides. Thus, it could be
  omitted in energy lanscapes, while it shows up in binding energies.

 The pairing interaction strengths $G$, Eq.~(\ref{bcs_ham}), are taken as
\begin{equation}\label{G_param}
G_l = (g_{0l}+g_{1l}I)/A ,
\end{equation}
 where the index $l$ stands for p (protons) or n (neutrons).

The strengths $G_l$ were fixed by adjusting
 the gap parameter $\Delta$ to the three-point odd-even mass differences
\begin{equation}
\begin{array}{l}
\Delta_Z M = (-1)^Z \left\{ \frac{1}{2}
\left[M(Z+1,N)+M(Z-1,N)\right] - M(Z,N) \right\}, \nonumber
\\[4mm]
\Delta_N M = (-1)^N \left\{ \frac{1}{2}
\left[M(Z,N+1)+M(Z,N-1)\right] - M(Z,N)\right\}. \vspace{3mm}
\end{array}
\end{equation}
The adjustment, using all measured masses of nuclei with
$Z$$\geqslant$88, resulted in the values \cite{mun01a}:
\begin{eqnarray}
g_{0l} = 17.67 \; {\rm MeV},  & g_{1l} = -13.11 \; {\rm MeV}, &
{\rm for} \; l={\rm n} \;{\rm (neutrons)}, \nonumber\\
g_{0l} = 13.40 \; {\rm MeV},  & g_{1l} =  44.89 \; {\rm MeV}, &
{\rm for} \; l={\rm p} \; {\rm (protons)}. \vspace{6mm}
\end{eqnarray}

\section{Shape parametrization}\label{sec.eqs}

 The essential point of any microscopic-macroscopic study is
 the kind and dimension of the deformation space
 used to describe a variety of nuclear shapes. This is particularly
 important for finding the saddle point along a fission path.
 Of course, there is no ideal shape parametrization.
 As far as we are interested in
 superheavy nuclei, with comparatively short fission barriers, a
 traditional expansion of the nuclear radius
 in spherical harmonics \cite{geom}, can be used.
 We admitt shapes of a 10D manifold defined by:
 \vspace{-2mm}
\begin{eqnarray}
 R(\vartheta ,\varphi)= R_0 c(\{\beta\})
\{ 1 &+&  \beta \left[ \cos\gamma
 {\rm Y}_{20}+\sin\gamma {\rm Y}_{22}^{(+)}\right]  \nonumber\\
 &+&\beta_{40}  {\rm Y}_{40} + \beta_{42} {\rm Y}_{42}^{(+)} +
\beta_{44}{\rm Y }_{44}^{(+)}\nonumber\\
  &+&\beta_{30}  {\rm Y}_{30} + \beta_{50}  {\rm Y}_{50} +\beta_{70}  {\rm Y}_{70}  \nonumber\\
  & + & \left. \qquad \beta_{60} {\rm Y}_{60} + \beta_{80} {\rm Y}_{80} \}
  \right].
 \end{eqnarray}
  The real spherical harmonics ${\rm Y}_{lm}^{(+)}$ are defined as:
  \begin{equation} {\rm Y}_{lm}^{(+)} = \frac{1}{\sqrt{2}} \left[ {\rm Y}_{lm} + (-1)^{m} {\rm Y}_{l - m}\right], \qquad {\rm for} \quad m \neq 0.
  \end{equation}
We use the conventional notation:
\begin{eqnarray}
\beta_{20}& = &\beta\cos{\gamma}, \nonumber\\
\beta_{22}& = & \beta\sin{\gamma},
\end{eqnarray}
where $\gamma$ is the Bohr quadrupole non-axiality parameter.
The function $c(\{\beta\})$ is determined by the volume-conservation condition.

 There is no physical principle which would forbid
 nonaxial ground-state nuclear shapes. However, calculations
 by M\"{o}ller et al. \cite{Mollerprl06} and our studies \cite{Jachim9,Jachim10,Jachim11}
 suggest that in the investigated nuclei the effect of nonaxiality (including
 octupole $Y^+_{32}$) in ground states is either small or non-existent.
 On the other hand, competing axially symmetric minima are frequent
 \cite{Jachim210}. Therefore, we assumed here
 the axial symmetry of the ground states. The energy is minimized
 simultaneously in all axial degrees of freedom:
$\beta_{20},\beta_{30},\beta_{40},\beta_{50},\beta_{60},\beta_{70},\beta_{80}$, using a
multidimensional conjugate gradient method.

 The saddle point is defined as a {\it minimum} over all paths connecting
 the ground state with the behind-the barrier region of the {\it maximal}
 energies along each path. Practical calculations are performed as follows.
 Energy is calculated at the following grid points
 (with steps given in parentheses):
\vspace{-5mm}
\begin{center}
\begin{eqnarray} \beta\cos\gamma& = &0(0.05)0.65, \nonumber\\
 \beta\sin\gamma& = &0(0.05)0.40, \nonumber\\
         \beta_{40}    & = &-0.20(0.05)0.20.
\end{eqnarray}
\end{center}
 Then, energy is
interpolated (by the standard SPLIN3 procedure of the IMSL
library) on the grid five times denser in each direction. Thus, we
finally have energy values at a total of $110946$ grid points.
 In order to find the saddle point a two-step
method is used. First, on such a 3-dimensional grid ($\beta_{20},
\beta_{22}, \beta_{40}$), the saddle point is determined by the Dynamic Programming
 Method given in \cite{Bellman} and adopted to the fission process by Baran et al. \cite{BARAN81}.
 Then, with these three deformations fixed,
 energy is minimized with respect to the other degrees of freedom:
$\beta_{42},\beta_{44},\beta_{30},\beta_{50},\beta_{60},\beta_{70},\beta_{80}$.
In the previous calculations \cite{CWIOK92,GHERGH99}, the nonaxial
hexadecapole deformations have usually been treated as functions
of the quadrupole triaxiality angle $\gamma$. In the present
calculations, the hexadecapole nonaxialities $\beta_{42}$ and
 $\beta_{44}$ are independent variables.

 \section{Dynamic programming method}

It is always possible to convert an m-dimensional grid:
($n_{1}\times n_{2}\times n_{3}\times ... \times n_{m}$) into a
four dimensional grid: ($n_{1}\times n_{2} \times n_{3}\times N$),
$N=n_{4} \times n_{5} \times n_{6} \times ... \times n_{m}$. In
the case of our deformation space: $n_{1}$ refers to $\beta_{20}$
(elongation), $n_{2}$ to $\gamma$ (nonaxility), $n_{3}$ to
$\beta_{40}$ (neck). The
 $N$ - axis describes all other degrees of freedom which we use for the
description of shapes (all other multipolarities). Each path $i$,
connecting the starting point with a behind-ther-barrier point
 $n+1$, may be characterized by the maximal value of
 energy $E_{max}^{i}$ which one can
 find along it, where $i$ is the index of a given path. The energy values
 between two neighboring points on a given path are
investigated with the help of an interpolation procedure. In this
way, we have a set of all possible paths $i$, connecting the
starting point to the $n+1$-th point, with the value of the
maximal energy $E_{max}^{i}$ on each. It is obvious that the
saddle-point energy will be the minimal value of all $E_{max}^{i}$
  over all possible paths (all possible $i$). The trajectory corresponding to this
 minimal value will automatically pass through the saddle point.
 It appears that to find the right trajectory along  which $E_{max}^{i}$ is
 minimal we do not need to consider all possible trajectories.

\section{Numerical tests and and error checks}

 The important numerical tool exploited here is the minimization procedure.
 It is used to find the ground state energy in a 7-dimensional space and
 the saddle point energy by the 7-dimensional minimization.
 Multidimensional minimization is a mixed blessing method: from the
 one point of view, it gives us the opportunity to find minima in
 the large deformation spaces (infeasible on a grid) but from the
 other, it introduces the necessity to check whether or not, the
 obtained minima are indeed the global ones.
 In order to gain some confidence in our results we used a number of checks.
 The standard checks within the minimization routine include the
monitoring of energy gradients. In addition, we looked at the
continuity of the resulting deformation parameters with respect to
$\beta\sin \gamma$ and $\beta\cos\gamma$ and at their stability
with respect to the choice of their starting values. The starting
values of the deformation parameters were always taken different
from zero.

 It should be also realized that we cannot be absolutely certain that
the minimization in the second step of our saddle-point-search procedure does
 not lead to errors. The hope that the initial deformation net
($\beta\sin\gamma,\beta\cos\gamma,\beta_{40}$) may be
sufficient is based mainly on the fact that other deformations
are small and weakly coupled to those three. In addition, we have checked
   saddle point energies obtained in the first stage of our procedure
  on the 3D grid by comparing them to the results of the analogous procedure
 using the 2D ($\beta\sin\gamma,\beta\cos\gamma$) and
 two variants of the 4D grids:
($\beta\sin\gamma,\beta\cos\gamma,\beta_{40},\beta_{42}$)
 and
($\beta\sin\gamma,\beta\cos\gamma,\beta_{40},\beta_{44}$).
 An important test of the saddle-point searching method was the
application of a completely different approach based on so-called
''{\it imaginary water flow}'' (IWF)
\cite{Luc91,Mam98,Hayes00,Moeler04,mol09}. This conceptually
simple method is still numerically efficient in the 5-dimensional
space
 of deformations $\beta\sin\gamma,\beta\cos\gamma,\beta_{40},\beta_{60},
 \beta_{80}$ and has been used for some nuclei. In order to avoid
 ambiguity, saddle points with vanishing parameters: $\beta_{42}, \beta_{44}$ were chosen.
  The obtained results were practically identical with those of the previous
method as one can see in Table \ref{test1}. One can see that the
difference in barrier does not exceed 110 keV.

\begin{table}
\caption{ 5D-IFW results } \label{test1}
%
\begin{tabular}{|c|c|c|c|c|c|c|c|c|}


 \hline
   Z      &   N     &    E [MeV]  &    $\beta_{20}^{sp}$ & $\beta_{22}^{sp}$ & $\beta_{40}^{sp}$ & $\beta_{60}^{sp}$& $\beta_{80}^{sp}$ \\
     \hline
    108      &   166     &    -1.05  &    0.33 & 0.01  & 0.04 & -0.01 & 0.05        \\
    112      &   174     &    -0.48  &    0.27 & 0.01  &-0.04 &  0.04 & 0.02        \\
   \hline
\end{tabular}


\end{table}


\section{Discussion of the results}

   The calculated properties of ground-states of even-even nuclei
 are given in Table 1 and those of saddle-points in Table 2.
 Energy maps in ($\beta \cos\gamma$, $\beta \sin\gamma$) plane, necessary
   to appreciate fission barriers, are shown in Fig. 1. They were obtained by minimization over 8 remaining
   deformations.
Total energy shown in Tables and Figures is normalized in such a
way that its macroscopic part is equal zero at the spherical
shape.

\begin{figure}[h!]

\begin{minipage}[h]{185mm}
 \includegraphics[width=0.6\linewidth,height=3in]{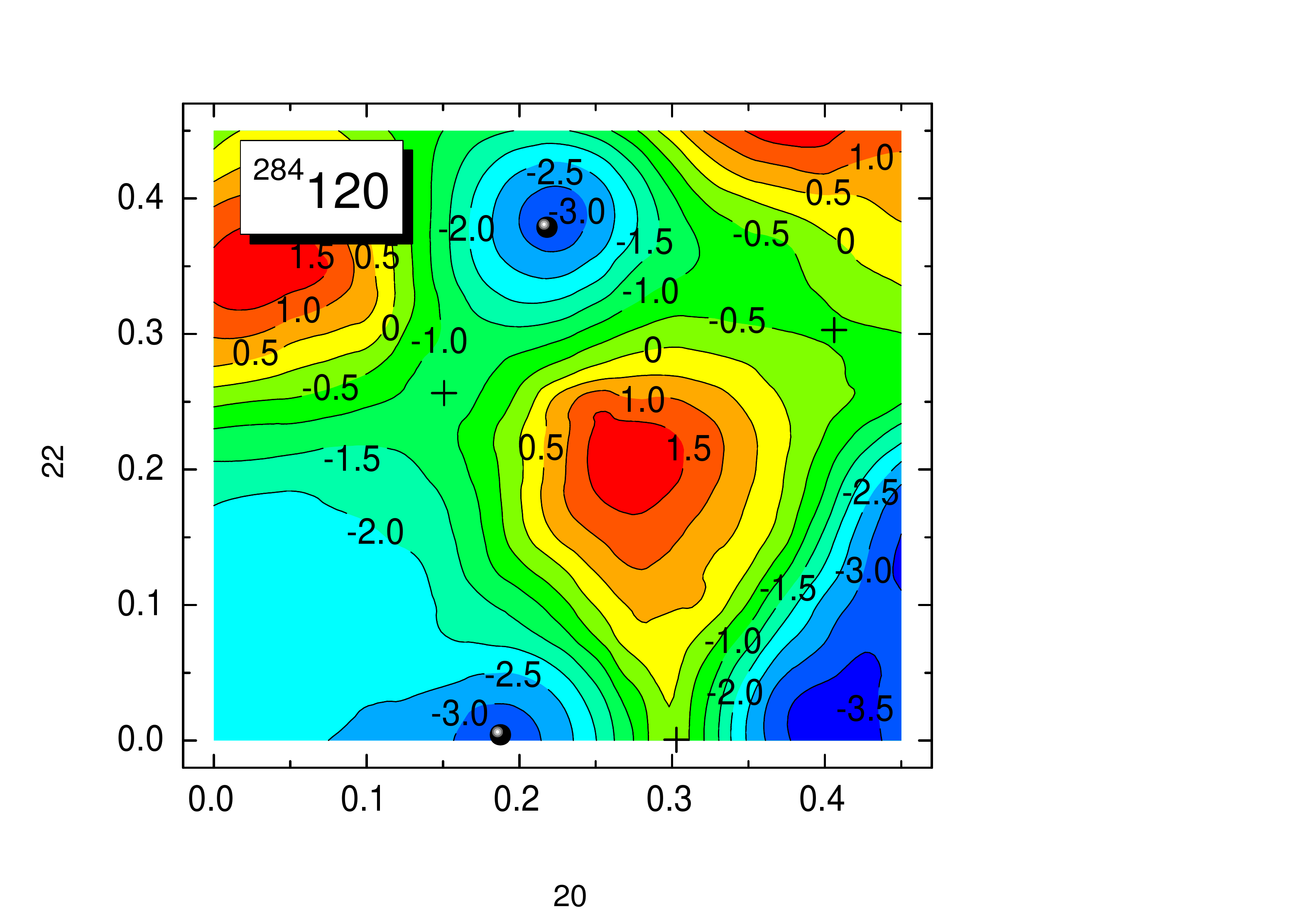}
 \includegraphics[width=0.6\linewidth,height=3in]{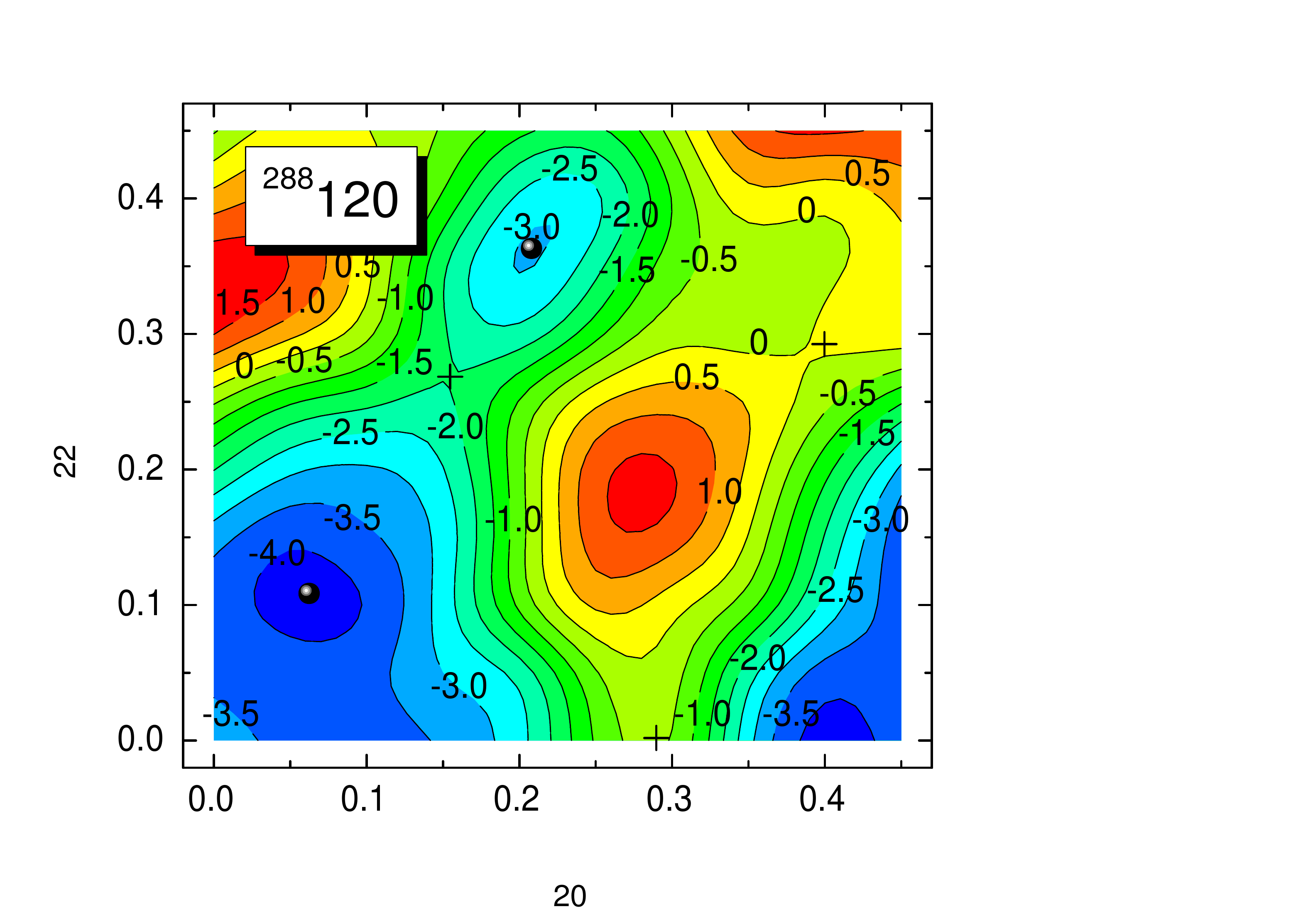}
  \end{minipage}
\begin{minipage}[h]{+185mm}
 \includegraphics[width=0.6\linewidth,height=3in]{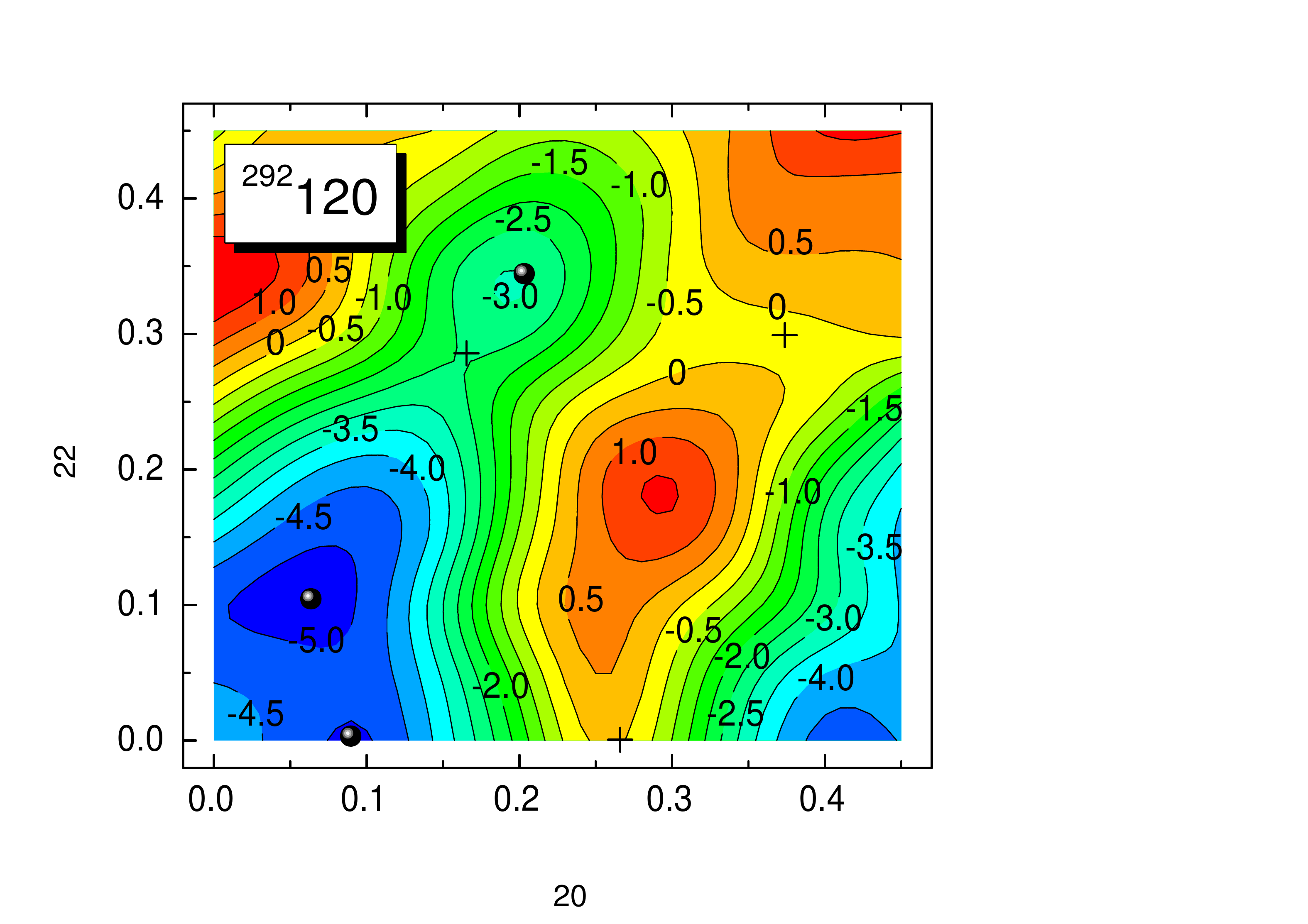}
 \includegraphics[width=0.6\linewidth,height=3in]{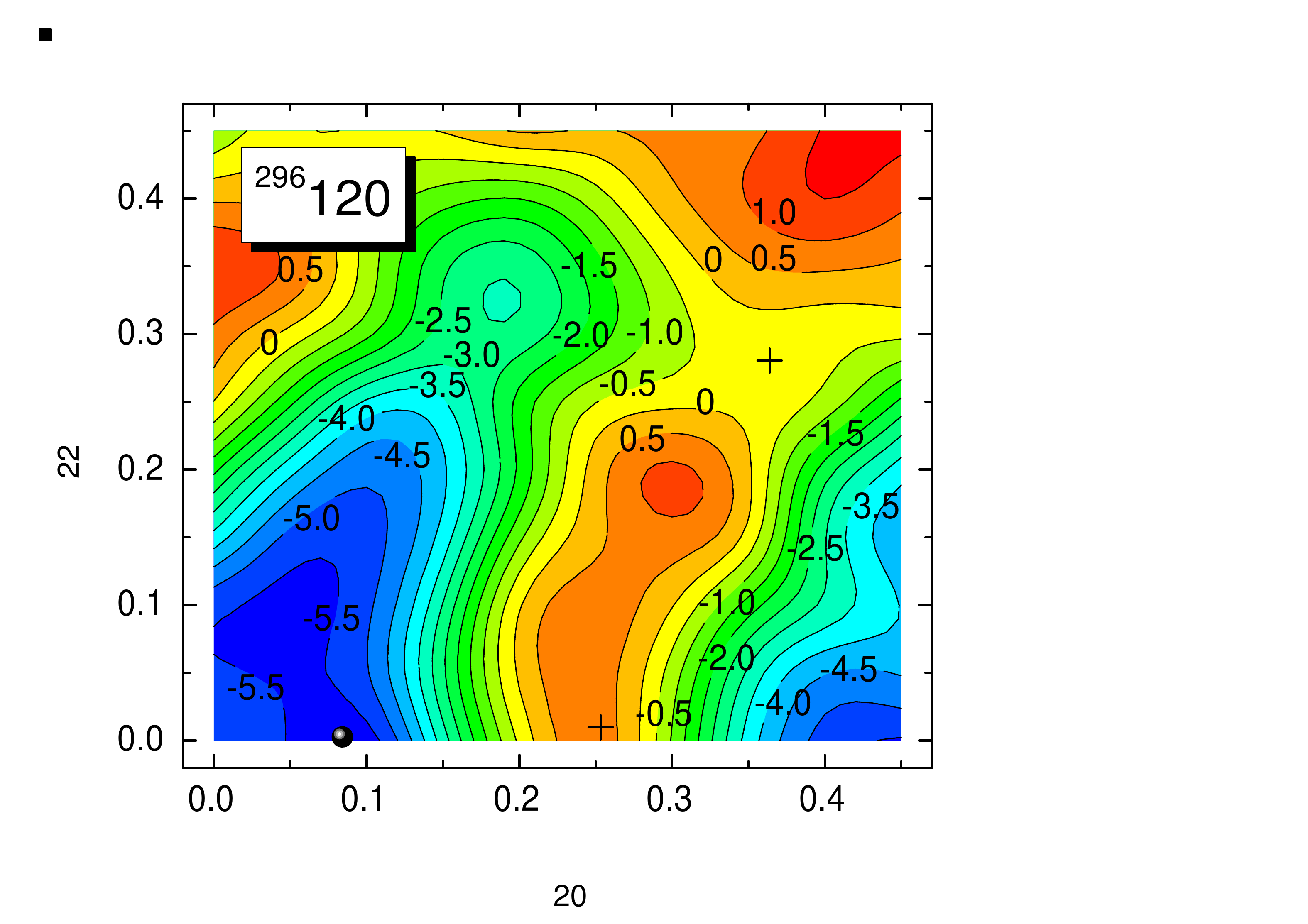}
  \end{minipage}
 \begin{minipage}[h]{+185mm}
 \includegraphics[width=0.6\linewidth,height=3in]{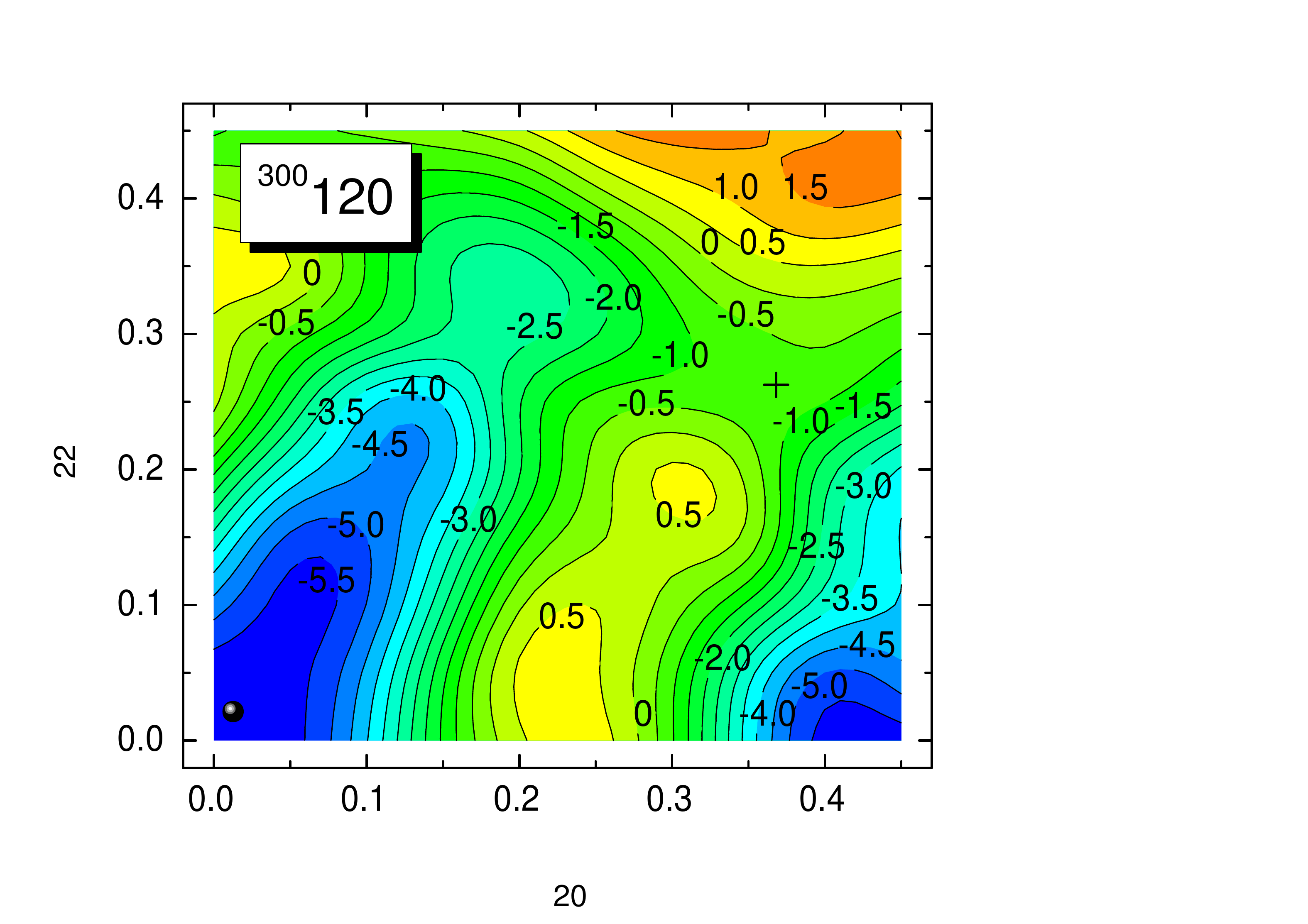}
 \includegraphics[width=0.6\linewidth,height=3in]{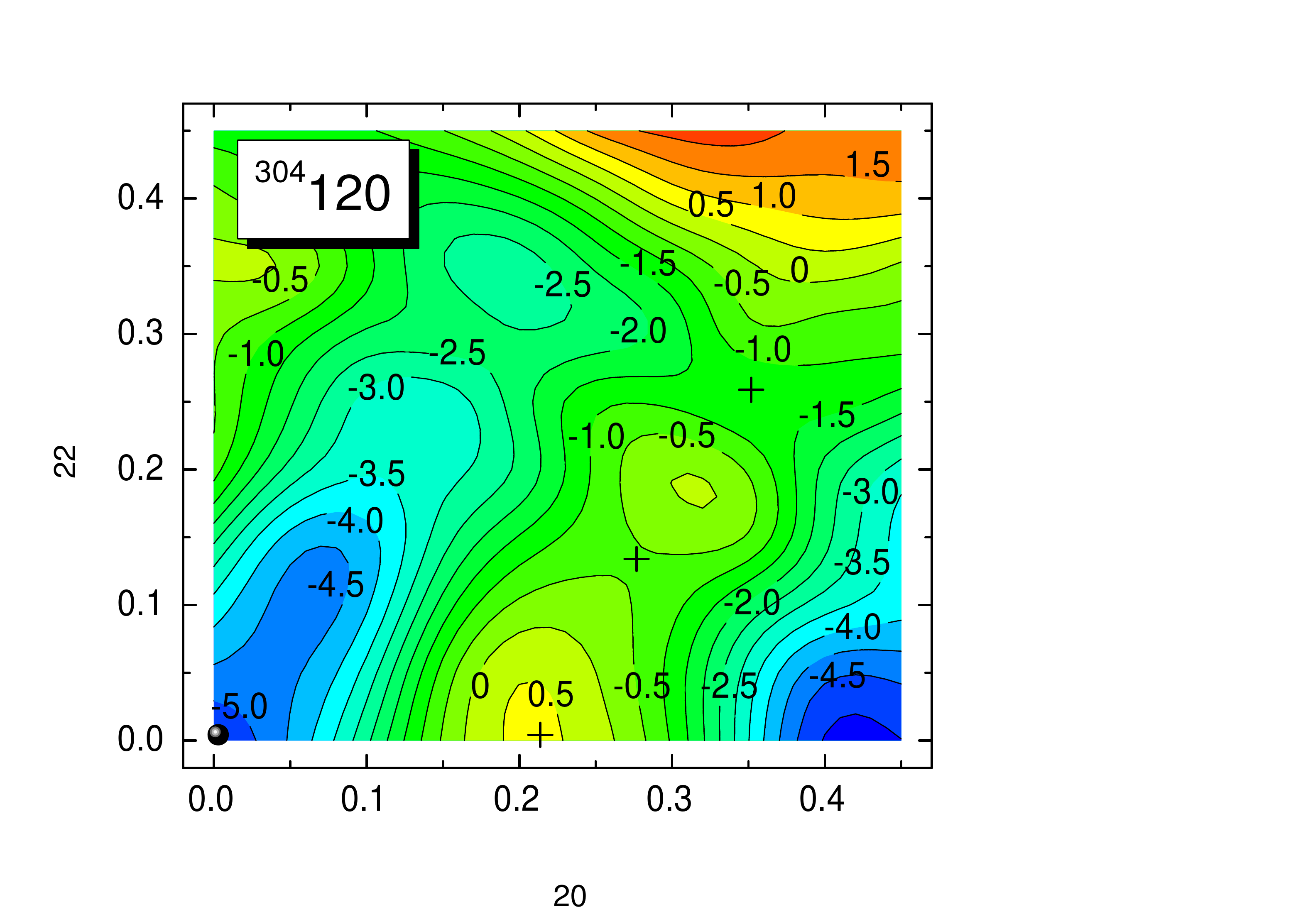}
 \end{minipage}
 \caption{Energy surfaces, $E-E_{mac}(sphere)$, for $Z=120$ isotopes,
       resulting from the minimization over the remaining 8 deformations.}
 \label{fig1}
\end{figure}

 \begin{figure}[h!]
  \begin{minipage}[h]{185mm}
  \includegraphics[width=0.6\linewidth,height=3in]{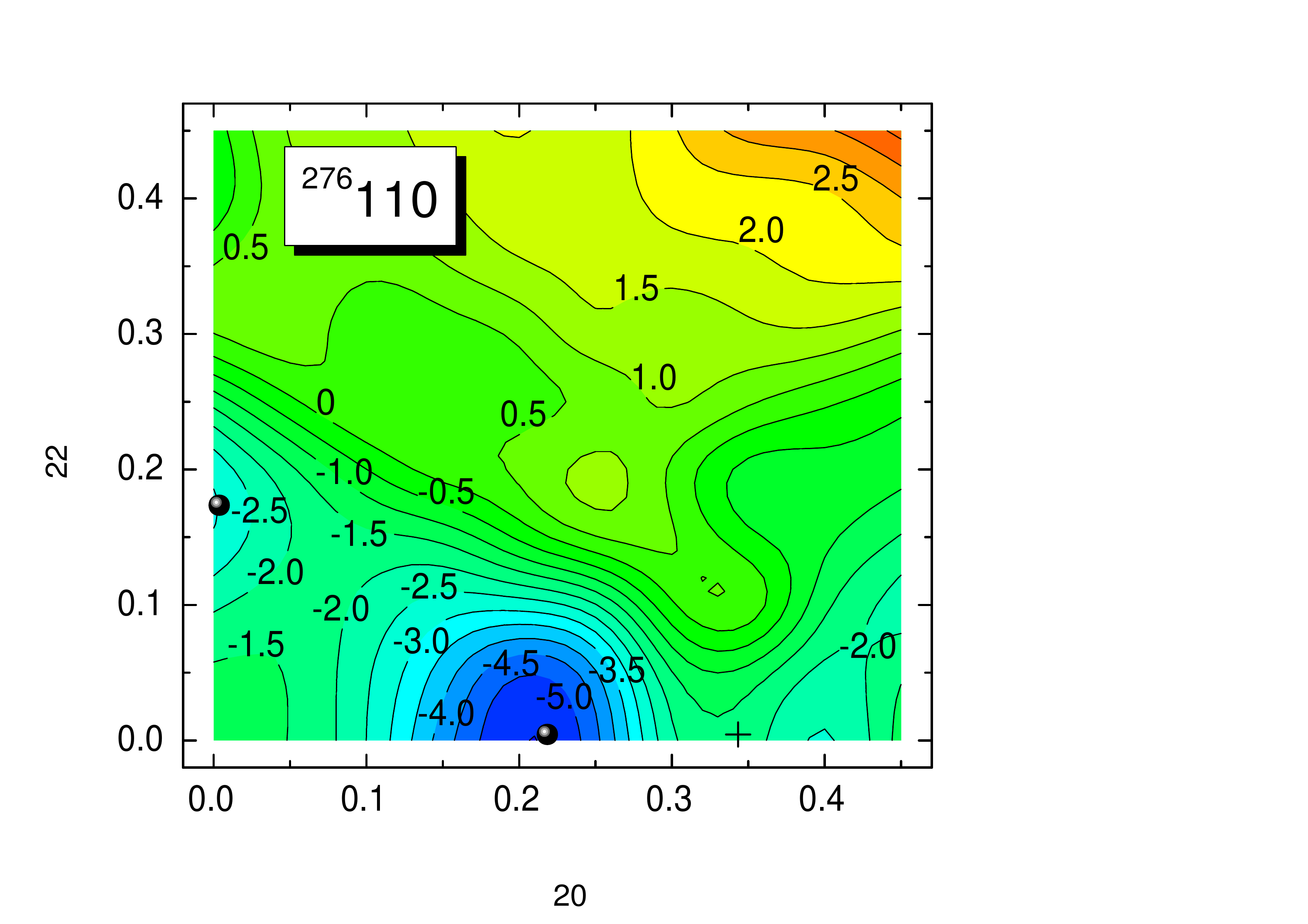}
  \includegraphics[width=0.6\linewidth,height=3in]{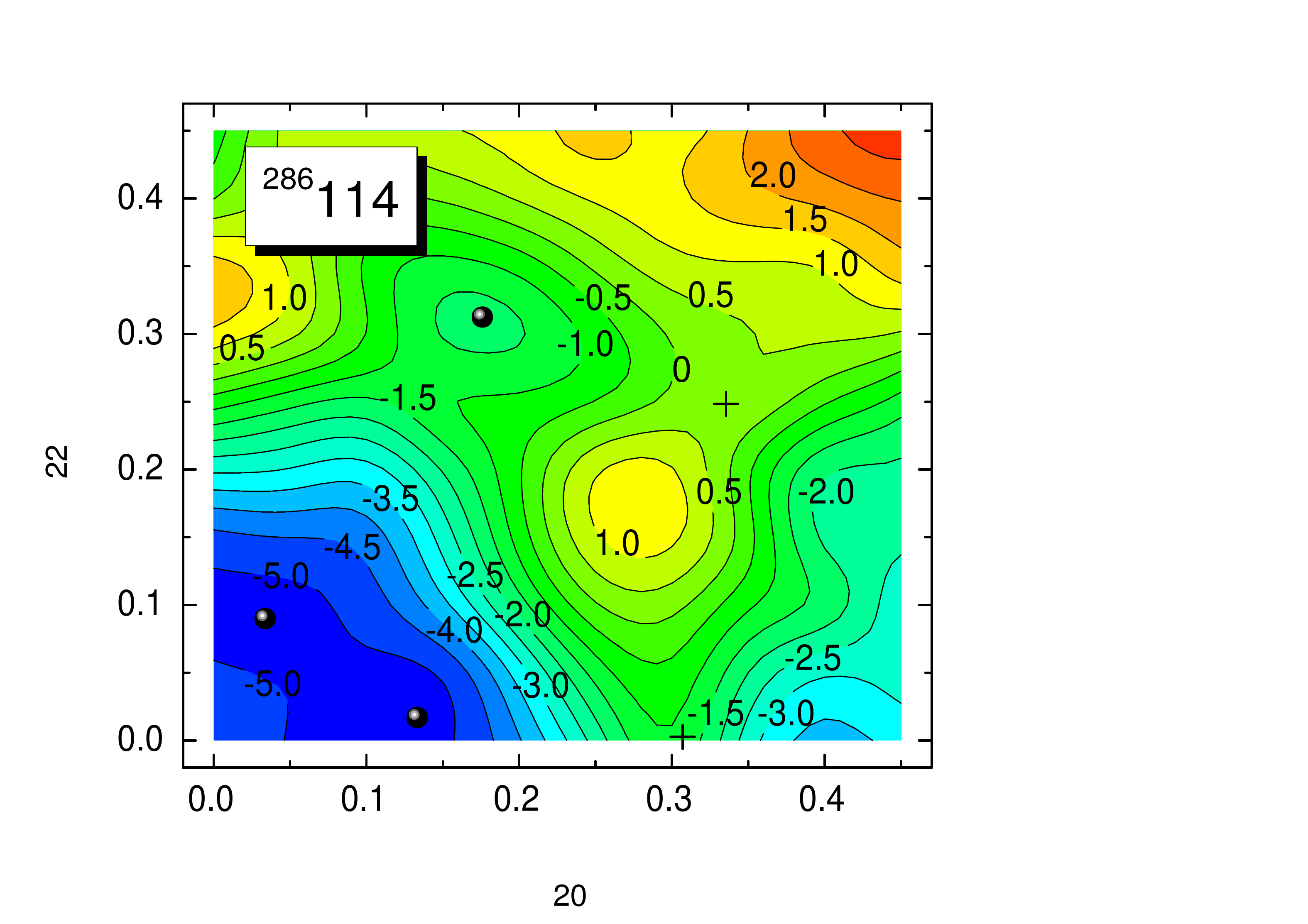}
  \end{minipage}
  \begin{minipage}[h]{185mm}
  \includegraphics[width=0.6\linewidth,height=3in]{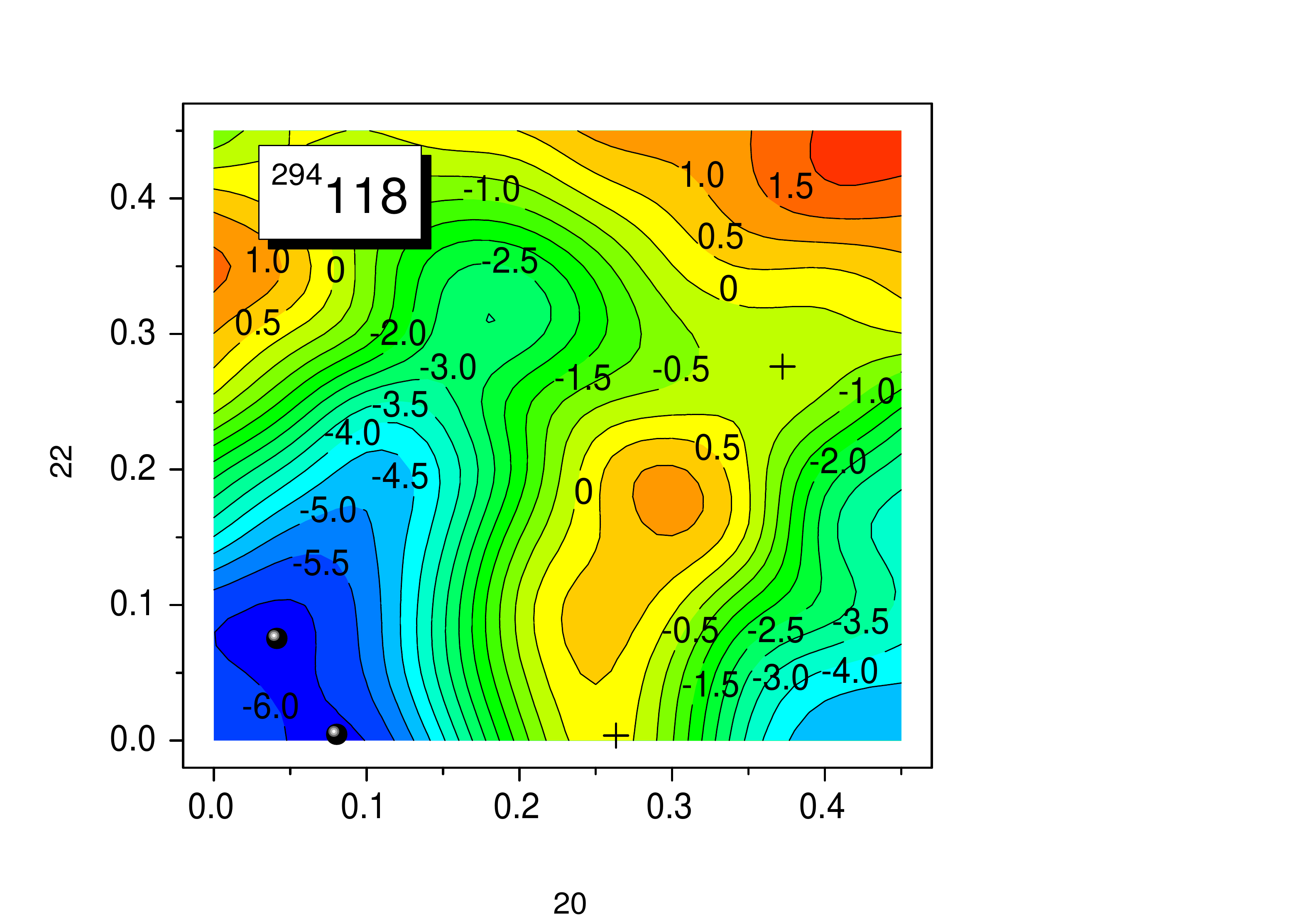}
  \includegraphics[width=0.6\linewidth,height=3in]{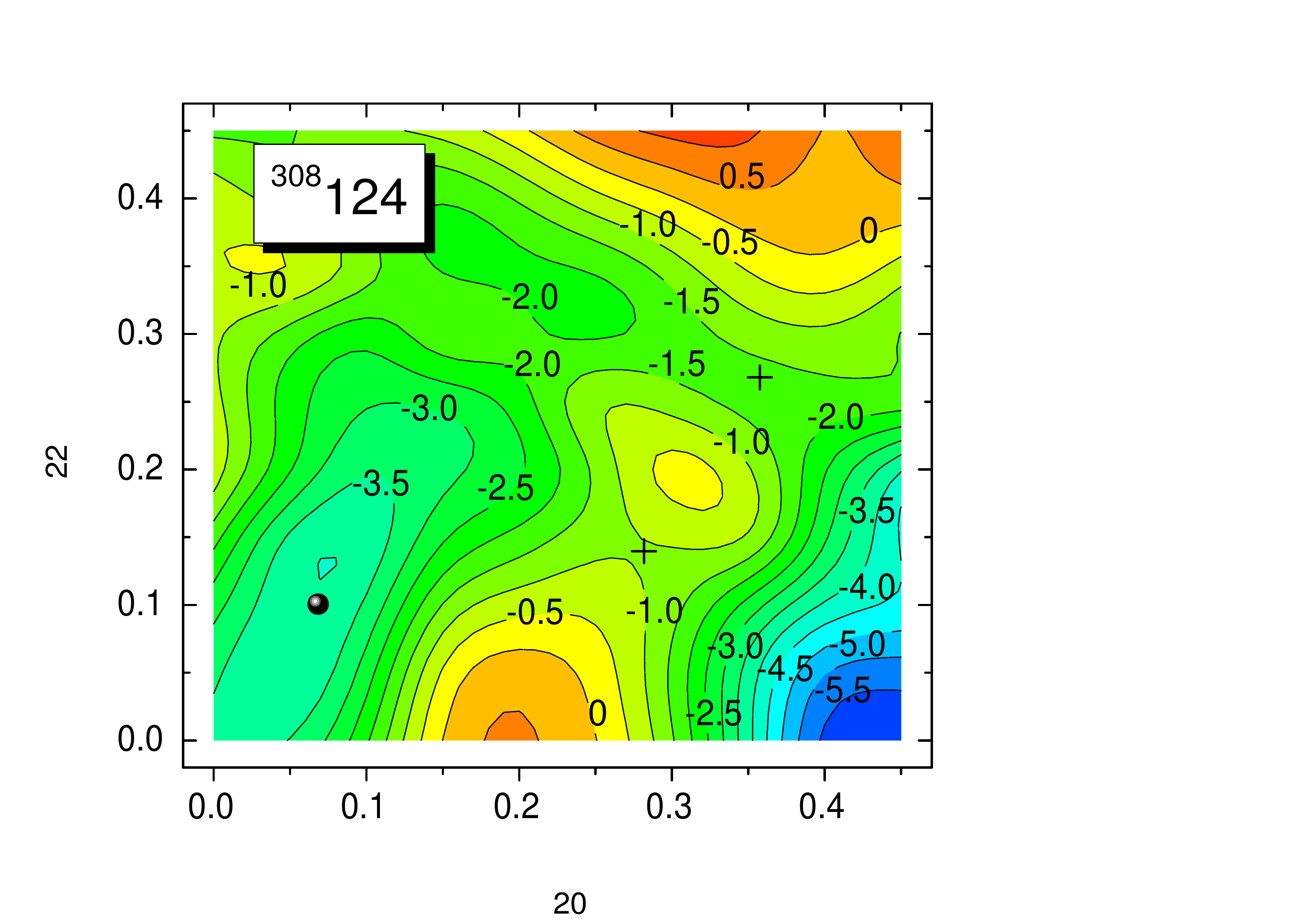}
  \end{minipage}
  \caption{As in Fig. 1, but for the three of experimentally detected
 evaporation residues and one hypothetical heavier system.}
  \label{fig2}
 \end{figure}

\subsection{Ground state properties}

\begin{figure}[ht!]
 \centering
 \includegraphics[width=0.8\linewidth,height=4in]{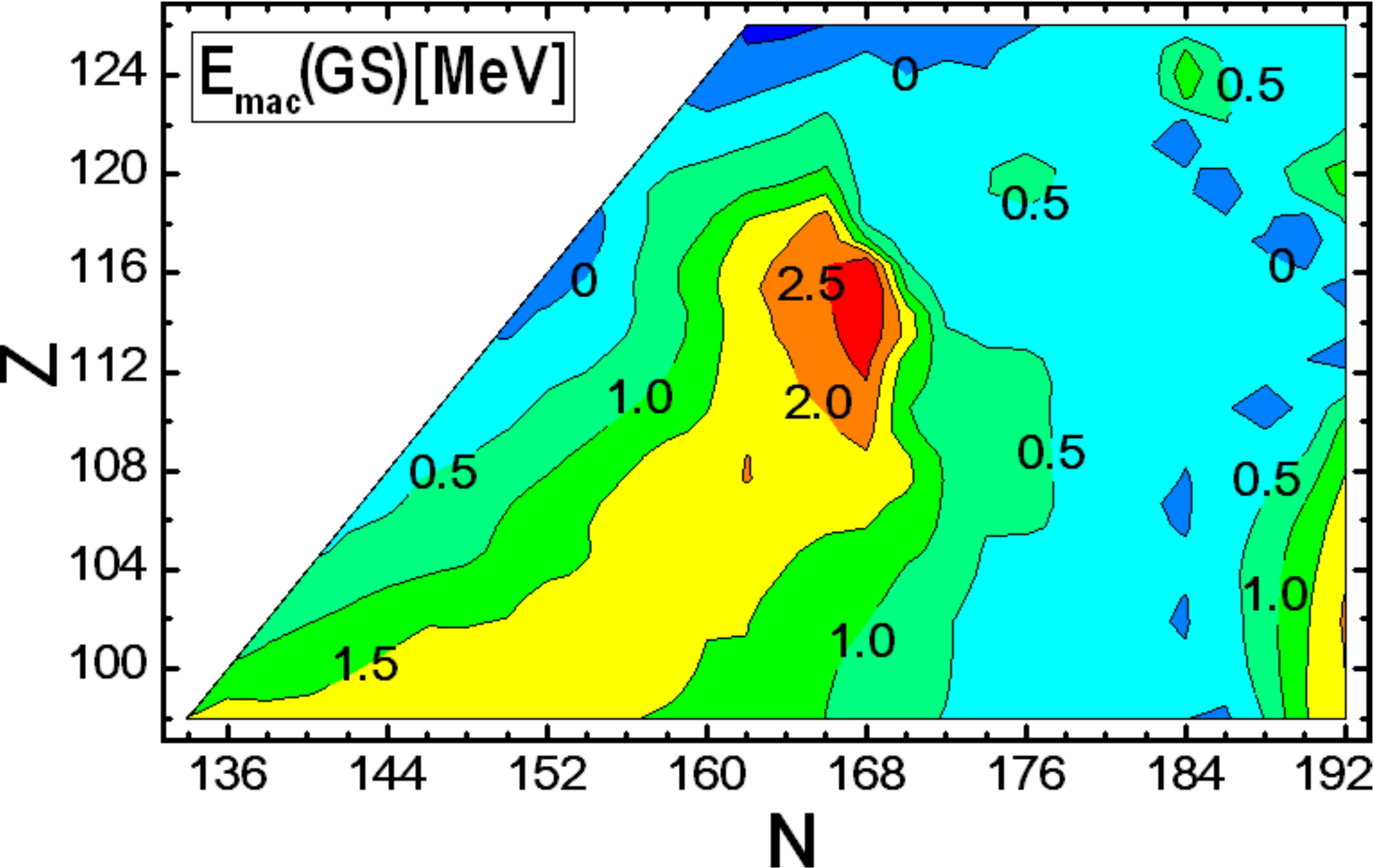}
  \includegraphics[width=0.8\linewidth,height=4in]{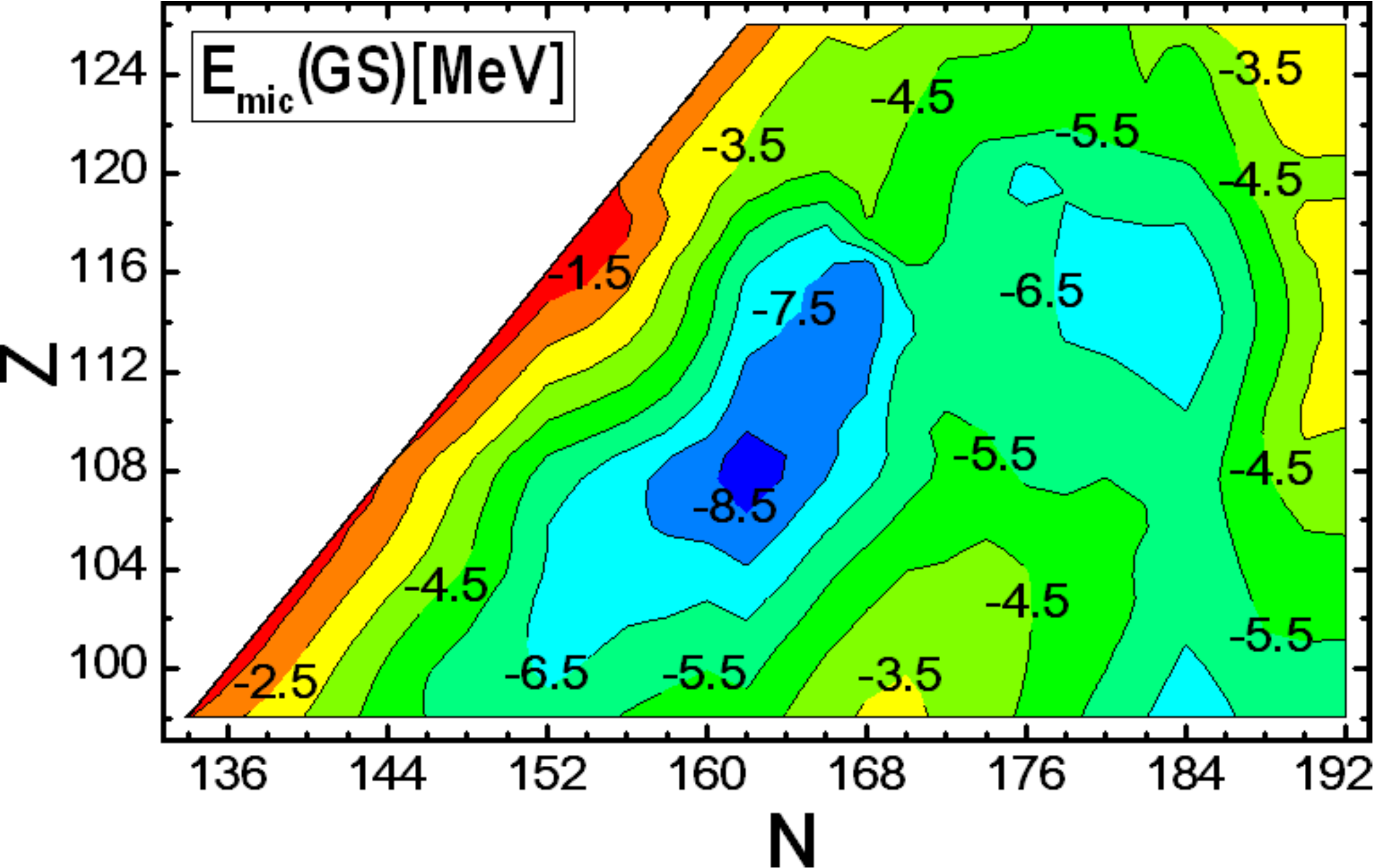}
 \caption{Calculated macroscopic and microscopic components of the
 ground-state binding energy, $E_{mac}-E_{mac}(sphere)$ and $E_{micr}$.}
 \label{fig3}
\end{figure}

To find the ground state masses and shapes the total energy is
minimized (using the gradient method) with respect to
$\beta_{20},\beta_{30},\beta_{40},\beta_{50},\beta_{60},\beta_{70},\beta_{80}$.
The candidates for the global minimum are chosen from the energy
map ($\beta_{20},\beta_{22}$) by inspection, taking into account
barriers heights. Only in a few cases we obtained a small nonzero
value of the octupole deformation and these are not given in
Tables. Macroscopic and microscopic parts of total energy are
shown in Fig. \ref{fig3}. The biggest shell effect ($\sim$ 9 MeV
)is observed for $^{270}$Hs
 ($Z=108$, $N=162$), the semi-magic nucleus.
As in other Woods-Saxon micro-macro calculations, the second minimum
 ($\sim$ 7MeV) of the shell correction
 is located around the nucleus $Z=114$ and $N=184$.
 When superposed with weakly deformation-dependent macroscopic part (Fig.\ref{fig3} ),
 this component is largely responsible for the emergence of global minima in
 superheavy nuclei.

\subsubsection{Ground state shapes (deformations)}

  \begin{figure}[ht!]
 \centering
 \includegraphics[width=0.8\linewidth,height=4in]{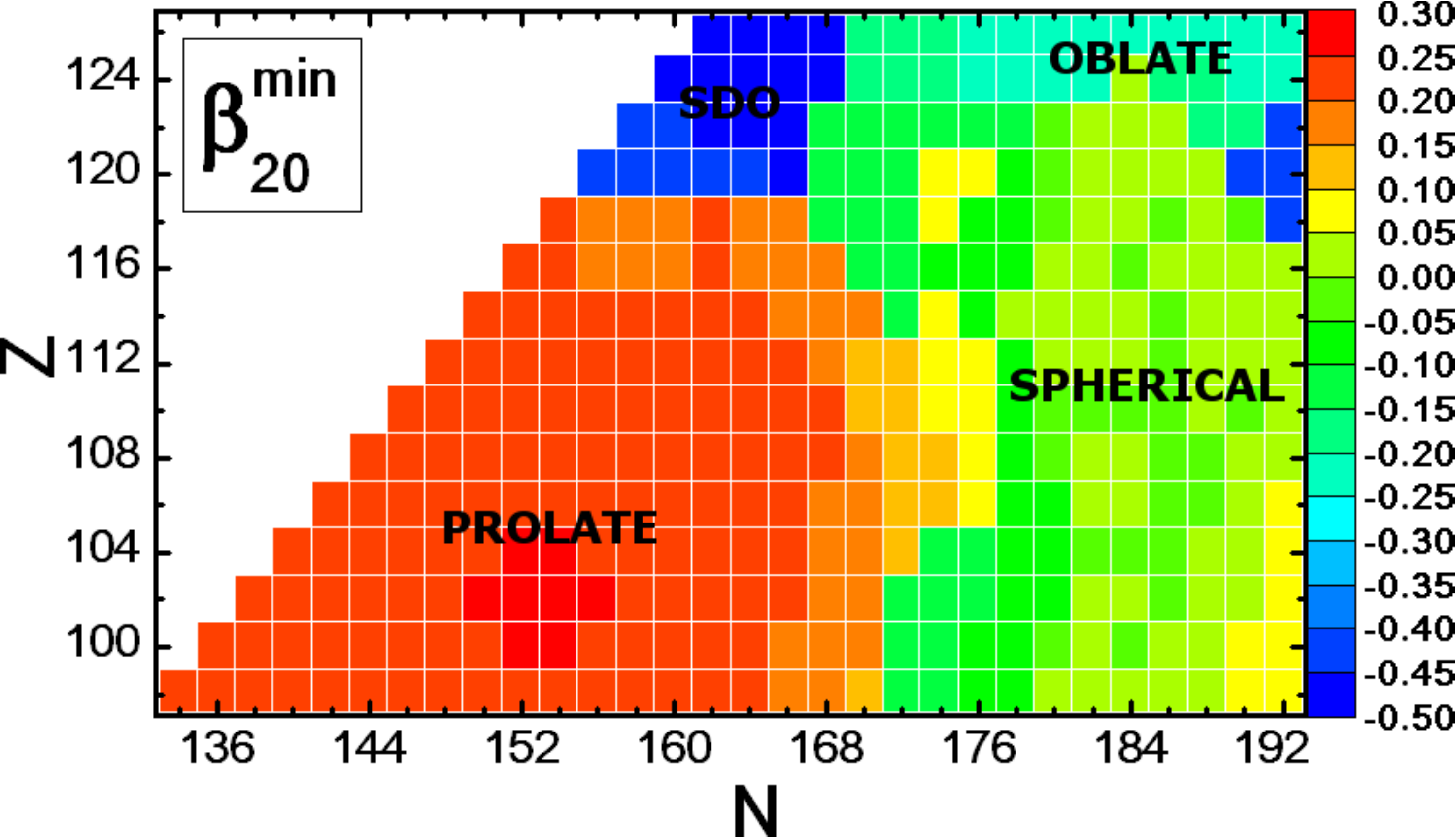}
 \caption{Calculated ground-state quadrupole deformations.}
 \label{fig6}
\end{figure}

  As mentioned in section 3, calculations including nonaxiall shapes show that the ground
  states are usually axially symmetric.
  For $Z>120$, there are a few minima exhibiting unususal shapes \cite{Jachim210},
  but they are at most degenerate with the minima given in Table 1.
   In addition to typical prolate, spherical and oblate shapes,
  the superdeformed oblate (SDO) shapes with $\beta_{20}\approx-0.45$ appear for
  some $Z\approx 120$ nuclei, discussed in the recent work \cite{SDO}.

   In the studied nuclei, up to $Z=118$, the shape evolution starts  with
  prolate shapes for small $N$ and ends with oblate or spherical (close to
 $N=184$) for largest $N$. The neutron-deficient $Z\geq 120$
  nuclei have SDO ground states. With increasing $N$, they evolve into
  oblate, then spherical, then oblate again, but for $Z=126$ all ground states
  are oblate. In some $Z\geq120$ systems, SDO ground states reappear for
  largest $N$ - see Fig. \ref{fig1}.

 Energy maps are necessary to study secondary minima and appreciate the
 competition of various shapes. The competition of prolate, spherical and
 oblate minima in $Z=120$ nuclei may be seen in Fig. \ref{fig1}. Let us note that
  the apparent secondary prolate minima at $\beta_{20}\approx 0.4$ are so shallow
  that they cannot be seriously considered as candidates for equilibrium
  configurations. In neutron deficient isotopes,
 the prolate and SDO shapes compete. With increasing $N$,
 the prolate minimum disappears and normal oblate minimum becomes lower
  than the SDO minimum. The spherical configuration becomes the g.s.
 for $N=180-184$. For still heavier isotopes (Table \ref{tab:A}, not shown in Fig.\ref{fig1}),
  the SDO minimum reappears as the g.s.

  In Fig. \ref{fig2}, energy landscapes are shown for three experimental nuclei (one
  synthetized in GSI and two in Dubna) and one hypothetical $Z=124$ system.
  The $^{276}$Ds is prolate, the two heavier, $^{286}114$ and $^{294}118$,
  are weakly deformed and $\gamma$-soft, while $^{308}124$ is weakly oblate.

\subsubsection{Ground state mass excess}

The accuracy of the approach may be assessed by comparing
 calculated and experimental masses (ground state masses)\cite{EXPMASS}.
The difference $M_{gs}^{th}-M_{gs}^{exp}$ is shown
 in Fig. 4  as a function of $A$ and in Fig. 5 as a function of both
  $Z$ and $N$.
 The accuracy is summarized in Table \ref{1} below.
The average discrepancy $<\mid M_{gs}^{th}- M_{gs}^{exp}\mid>$, the maximal difference
$Max \mid M_{gs}^{th}- M_{gs}^{exp}\mid$ and the $r.m.s.$ deviation are shown for a number
$N$ of even-even superheavy nuclei.

\begin{table}
\caption{ Statistical parameters of the calculated mass excess in
relation to experimental or recommended atomic mass excess
\cite{EXPMASS}.
 All quantities are in MeV, except for the number of nuclei N. } \label{1}
%
\begin{tabular}{|c|c|c|c|}


 \hline
   N      &   $<\mid M_{gs}^{th}- M_{gs}^{exp}\mid>$     &    $Max \mid M_{gs}^{th}- M_{gs}^{exp}\mid$  &    r.m.s  \\

     \hline

      67                                         & 0.43    &  1.58   &   0.58        \\

   \hline
\end{tabular}


\end{table}

\begin{figure}[h]
 \centering
 \includegraphics[width=0.8\linewidth,height=8in]{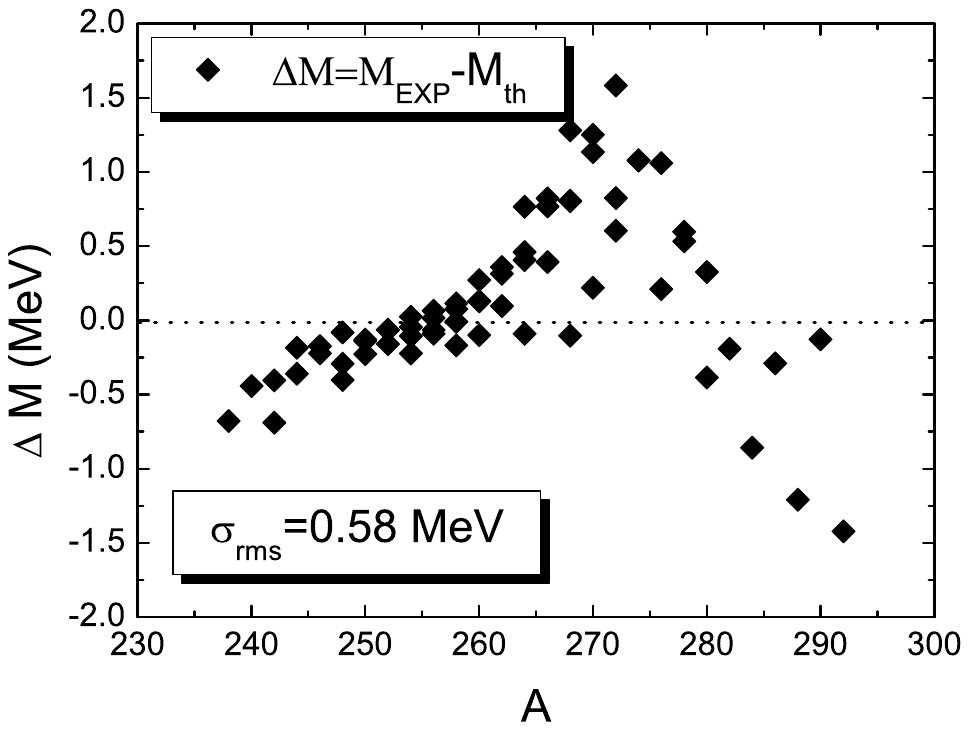}
 \caption{Discrepancy between the measured and calculated nuclear masses vs
 mass number.}
 \label{fig4}
\end{figure}

The agreement between the calculated and experimental masses is the worst for
 nuclei located near $Z=106$, $N=162$ ($\simeq$ + 1.5 MeV) and $Z=114$, $N=184$
  ($\simeq$ -1.5MeV).

  \begin{figure}[ht!]
 \centering
 \includegraphics[width=0.8\linewidth,height=5in]{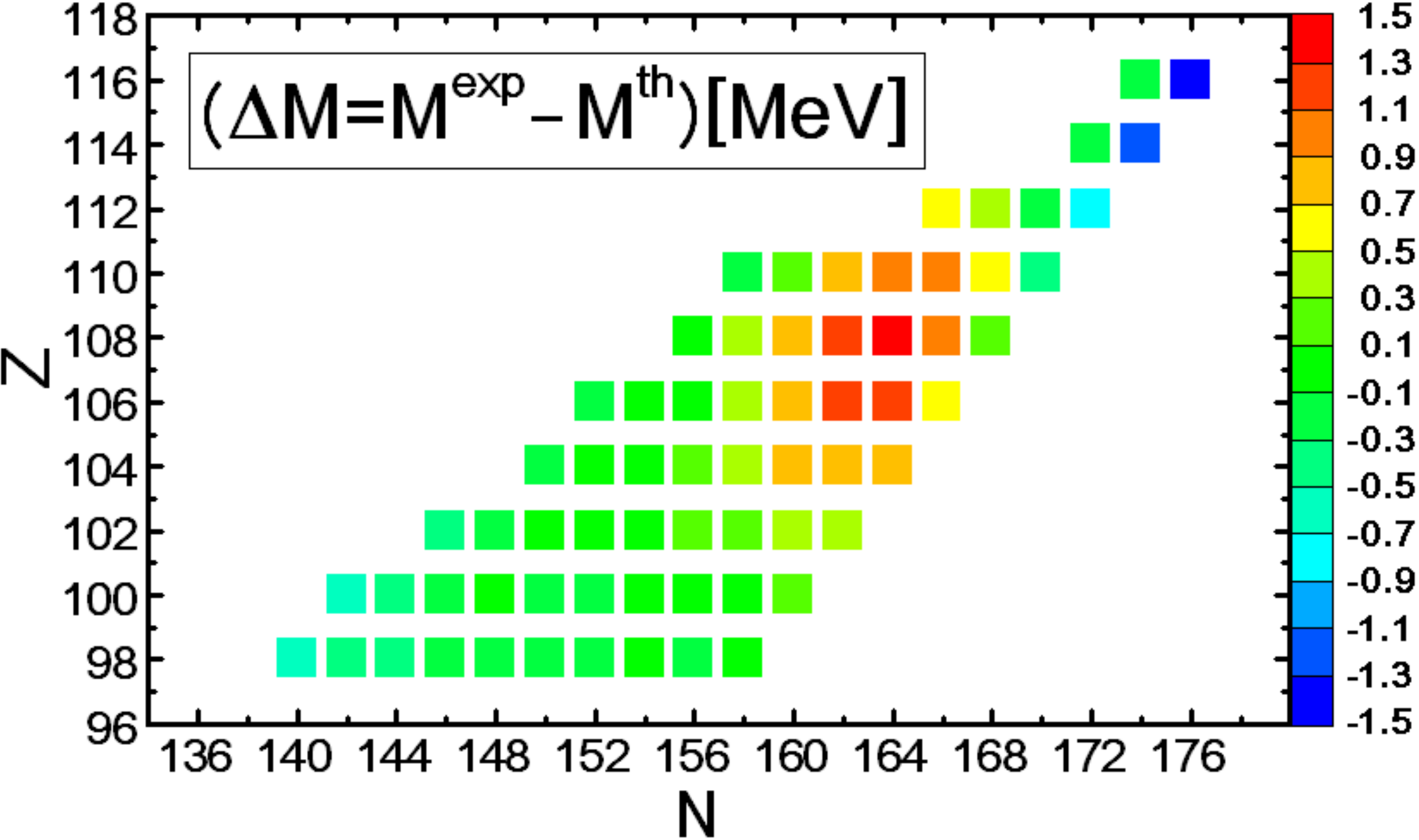}
 \caption{As in Fig. 4, but as a function of proton and neutron numbers.}
 \label{fig5}
\end{figure}



\subsubsection{Q-alpha energies}

  $Q^{th}_{\alpha}$ values given in Table 2 are calculated always for g.s. to g.s
  transitions, even if the corresponding deformations differ widely and
  hence one can expect a substantial decay hindrance. $Q_{\alpha}$
 energy for a nucleus with N neutrons and Z protons
 can be directly obtained from masses

\begin{equation}
Q_{\alpha}^{th} (Z,N)
=M_{gs}^{th}(Z,N)-M_{gs}^{th}(Z-2,N-2)-M(2,2).
\end{equation}

\begin{figure}[ht!]
 \centering
 \includegraphics[width=0.8\linewidth,height=4in]{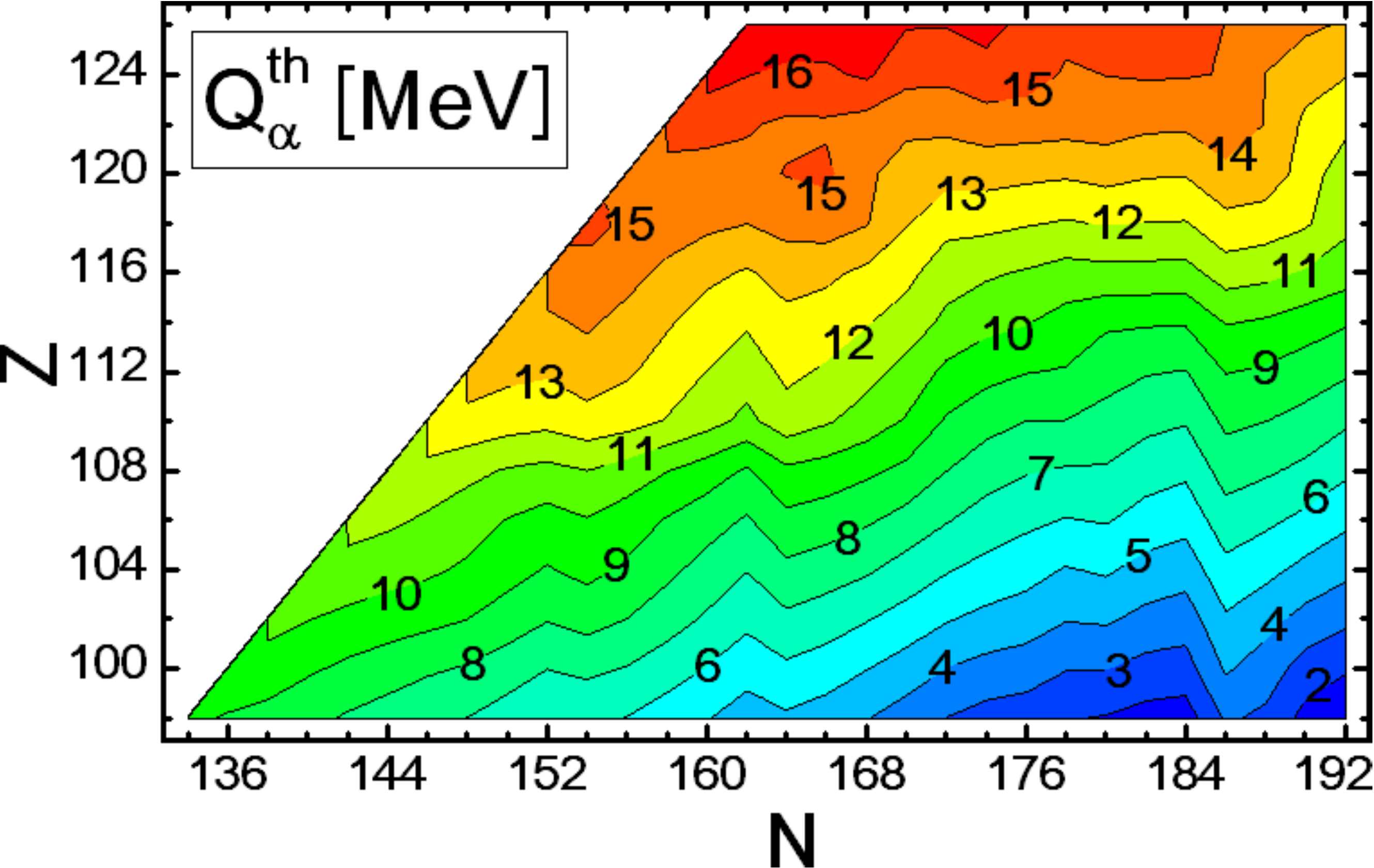}
 \caption{Calculated $\alpha$-decay energies for g.s.$\rightarrow$g.s
  transitions as a function of proton and neutron numbers.}
 \label{fig7}
\end{figure}

 The resulting discrepancies between
 calculated and experimental $Q_{\alpha}$ values are shown in Table
 \ref{2}.
The average discrepancy  $<\mid Q^{th}_{\alpha}- Q^{exp}_{\alpha}\mid>$ , the maximal difference
$Max \mid Q^{th}_{\alpha}- Q^{exp}_{\alpha}\mid$ and the $r.m.s.$ deviation are shown for a number
$N$ of even-even superheavy nuclei.

\begin{table}
\caption{ Statistical parameters of calculated $Q_{\alpha}$ energies in
relation to experimental data \cite{EXPMASS} . All quantities are in MeV, except for the
number of nuclei N.   \label{2}}

\begin{tabular}{|c|c|c|c|}


 \hline
   N      &   $<\mid Q^{th}_{\alpha}- Q^{exp}_{\alpha}\mid>$     &    $Max \mid Q^{th}_{\alpha}- Q^{exp}_{\alpha}\mid$ &    r.m.s  \\

     \hline

      67                                         & 0.02  &  0.74  &   0.29        \\

   \hline
\end{tabular}


\end{table}

 The largest discrepancy between the calculated and experimental $Q_{\alpha}$
 values results for nuclei located near $Z=106$, $N=162$,
  what is a consequence of the calculated masses in this area.

\begin{figure}[ht!]
 \centering
 \includegraphics[width=0.8\linewidth,height=4in]{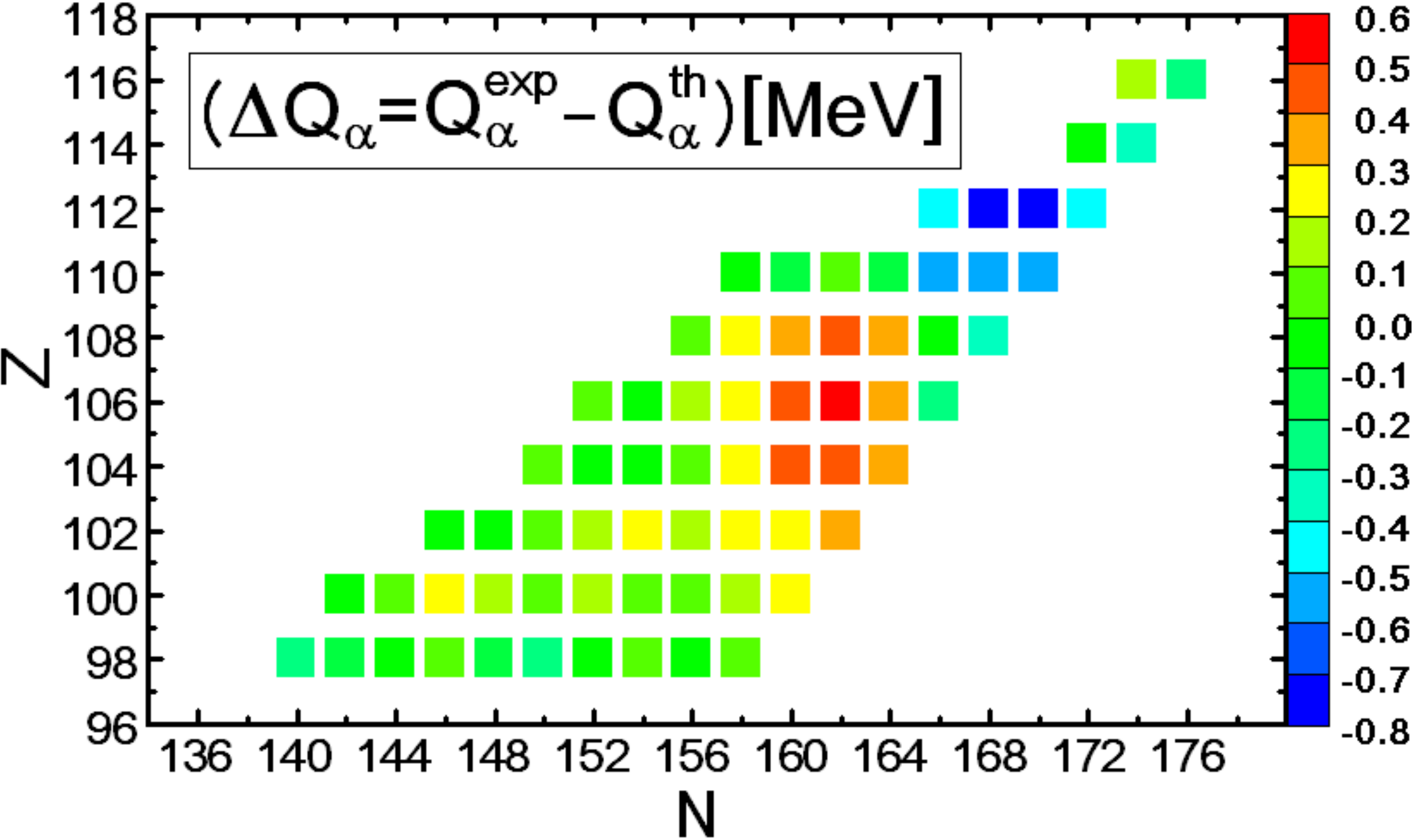}
 \caption{Discrepancy between the experimental and calculated $\alpha$-decay
 energies as a function of proton and neutron numbers.}
 \label{fig8}
\end{figure}

\subsection{Saddle point properties}

  It is worth emphasizing that the saddles listed in Table 2 correspond to
  the ground states. In case that secondary minima exist, their saddle
  points may be {\it different} from that of the ground state. In other words,
  a saddle point always relates to a minimum (equilibrium, metastable state)
  whose decay it characterizes.

\begin{figure}[ht!]
 \centering
 \includegraphics[width=0.8\linewidth,height=4in]{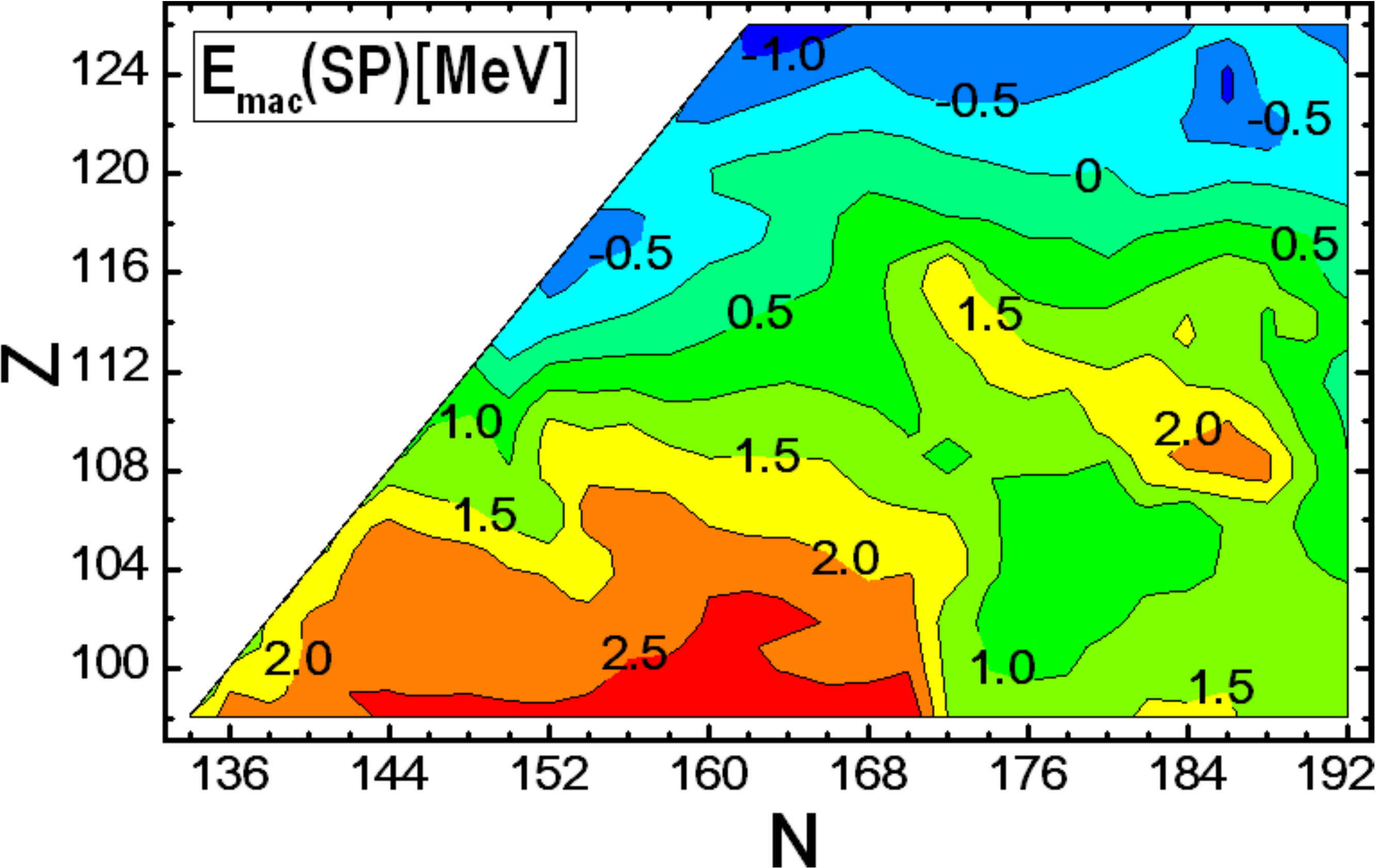}
  \includegraphics[width=0.8\linewidth,height=4in]{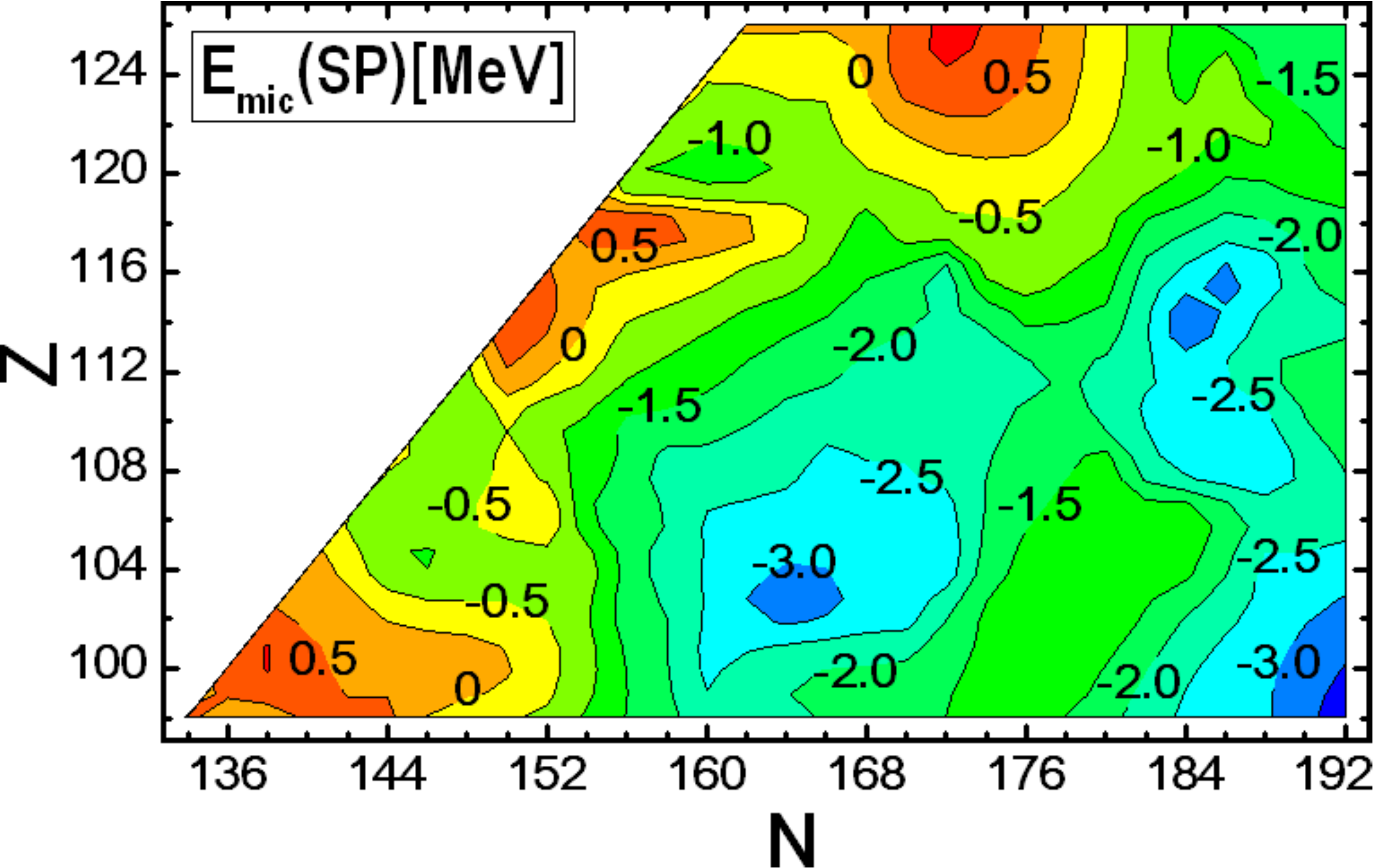}
 \caption{As in Fig. 3, but for the calculated saddle points.}
 \label{fig9}
\end{figure}

  The used method of the search for saddles, the combined 3D saddle-point
  search and subsequent 7D minimization, implies that they may slightly
 differ from those read from the maps in ($\beta\cos\gamma, \beta\sin\gamma$),
 shown in Figs. 2 and 3. The latter are fixed by the single 8D minimization for
   the purpose of illustration of an energy landscape.
      For example, the axially-symmetric saddle in $^{288}120$ seems to be
     lower than the triaxial one in Fig. 2, while it is the opposite in Table 2.
       The local shell correction and
 the macroscopic energy at the saddle point configuration has been shown in Fig. \ref{fig9}.
Generally, we see that the shell effects at the saddle point are much weaker than in the ground state,
 but not negligible. The most pronounced effects ($\sim$ 3MeV) were obtained around nuclei
with Z=102, N=162 and Z=114, N=184. For the majority of nuclei,
microscopic energy at the saddle point is negative while
macroscopic energy is positive. Values of these energies are
similar in magnitude.

 \subsubsection{Saddle point shapes (deformations)}

   The calculated saddles are mostly triaxial in $Z=98-104$, and exclusively
  triaxial in $Z\geq 122$ nuclei.
  There are many triaxial saddles with rather sizable values of the deformation
  $\gamma$. Since the saddle deformation is a result of a competition
  between the axial and triaxial saddle energies, even their tiny difference may
  result in an abrupt change in $\gamma^{sp}$.  This is reflected in the abrupt
 changes of $\gamma^{sp}$ between neighbouring isotopes - Table 2.
  Stated otherwise, a large $\gamma^{sp}$ does not necessarily correlate with
 a large nonaxiality effect on the saddle-point energy (mass).

 \subsubsection{Saddle point mass excess - fission barriers}

  Fission barriers may be obtained from Tables 1 and 2 as:
 \begin{equation}
B_{f}=  E^{th}_{sp}-E^{th}_{gs}=M^{th}_{sp}-M^{th}_{gs}.
\end{equation}
Most of them were given and discussed in \cite{BAR10}. One can
compare our barriers heights with other recent calculations
\cite{DOB,mol09,Abusara10,Abusara12,Erler} .
  They are relevant for fission rates from excited states (thermal rates),
  with the excitation energy greater than, roughly, the barrier itself.
  The spontaneous fission rates involve an additional inertia effect,
  see e.g. \cite{Smol}.

 \section{Summary}

Using the macroscopic-microscopic model, we have calculated the
ground state and saddle point properties of even-even superheavy
elements: masses, macroscopic energies, shell corrections,
deformations as well as the ground state to ground state alpha
decay energies.

\ack This work was supported  by the Polish Ministry of Science
and Higher Education, Contract No. JP201 0013570.
 One of the authors (P.J.) was
co-financed by the European Social Fund and the state budget
(within Sub-measure 8.2.2 Regional Innovation Strategies, Measure
8.2 Transfer of knowledge, Priority VIII Regional human resources
for the economy Human Capital Operational Programme).

\section*{Table 1. Ground state properties}

For the isotopes of the elements $Z$=98-126, tabulates the ground
state masses, total energies, macroscopic and microscopic
energies, equilibrium deformations and corresponding
$\alpha$-decay ground state to ground state $Q$-values.

\begin{center}


\section*{Table 2. Saddle point properties}

For the isotopes of the elements $Z$=98-126, tabulates the saddle
point masses, total energies, macroscopic and  microscopic
energies as well as saddle point deformations.

\begin{center}
\begin{tabular}{lllllll}
$Z$&      The atomic number\\
$A$&      The mass number\\
$^{th}_{sp}$&   The mass excess  in MeV\\
$E$& The total energy in MeV \\
$E_{mac}$& The macroscopic energy in MeV \\
$E_{mic}$&  The microscopic energy  in MeV\\
$\beta^{sp}_{20} \div \beta^{sp}_{80}$&  The saddle point deformations\\
$Q_{alpha}$& Theoretical $Q$-value in MeV \\
\end{tabular}
\end{center}
\clearpage
\datatables 
\LTright=0pt \LTleft=0pt

\footnotesize

\begin{longtable}{@{\extracolsep\fill}rrrrrrrrrrrrr}
 \caption{Calculated Saddle Point Masses and Deformations.
\label{tab:B}}\\[3pt]
\hline \hline\noalign{\vskip3pt}
  Z & N& $M^{th}_{sp}$& E & $E_{mac}$& $E_{mic}$ & $\beta^{sp}_{20}$ & $\gamma^{sp} (^{\circ})$ & $\beta^{sp}_{40}$
  & $\beta^{sp}_{42}$& $\beta^{sp}_{44}$ & $\beta^{sp}_{60}$ &$\beta^{sp}_{80}$    \\[3pt]

  \hline\noalign{\vskip3pt}
\endfirsthead
 \multicolumn{6}{@{}c@{}}{Table  \ref{tab:B} contd\dots}\\
 \hline \hline\noalign{\vskip3pt}
  Z & N& $M^{th}_{sp}$& E & $E_{mac}$& $E_{mic}$ & $\beta^{sp}_{20}$ & $\gamma^{sp} (^{\circ})$ & $\beta^{sp}_{40}$
  & $\beta^{sp}_{42}$& $\beta^{sp}_{44}$ & $\beta^{sp}_{60}$ &$\beta^{sp}_{80}$    \\[3pt]

  \hline\noalign{\vskip3pt}
\endhead
 \hline
 \noalign{\vskip3pt}
 \multicolumn{6}{r}{\itshape Continued\dots}\\
\endfoot
 \noalign{\vskip3pt}\hline
\endlastfoot
98  & 134 &  61.77 &  2.33 &  1.64 &  0.68 &   0.393   &    7.3    &  -0.010   &  0.009 & -0.003 & -0.014 & -0.019  \\
98  & 136 &  61.12 &  2.12 &  2.18 & -0.06 &   0.412   &    5.6    &  -0.030   &  0.015 &  0.003 & -0.022 & -0.015  \\
98  & 138 &  61.49 &  2.35 &  2.18 &  0.18 &   0.421   &    2.7    &  -0.030   &  0.008 &  0.003 & -0.026 & -0.008  \\
98  & 140 &  62.69 &  2.82 &  2.23 &  0.59 &   0.440   &    0.0    &  -0.030   &  0.000 & -0.002 & -0.025 &  0.001  \\
98  & 142 &  64.26 &  3.10 &  2.45 &  0.65 &   0.475   &    8.5    &  -0.010   &  0.017 &  0.022 & -0.011 &  0.001  \\
98  & 144 &  66.18 &  3.15 &  2.55 &  0.61 &   0.510   &   11.3    &   0.000   &  0.000 &  0.023 &  0.006 & -0.002  \\
98  & 146 &  68.18 &  2.75 &  2.85 & -0.10 &   0.526   &   14.3    &   0.030   &  0.002 &  0.025 &  0.020 &  0.005  \\
98  & 148 &  70.79 &  2.43 &  2.91 & -0.48 &   0.526   &   14.3    &   0.030   &  0.003 &  0.025 &  0.018 &  0.007  \\
\end{longtable}

* marginally unbound nucleus

\clearpage

\end{document}